\documentclass[iop,twocolappendix,numberedappendix]{emulateapj}
\usepackage{epsfig}
\usepackage{graphics,graphicx}
\usepackage{tikz}
\usepackage{subfigure}
\usepackage{multirow}
\usepackage{amssymb}
\usepackage{natbib}
\usepackage{rotating}
\usepackage{setspace}
\usepackage[bookmarks=false, colorlinks=true, citecolor=blue, backref=true]{hyperref}
\usepackage{apjfonts}

\shorttitle{ALMA Observations of NGC~1614}
\shortauthors{C.K. Xu et al.}

\newcommand{\lsim}{\, \lower2truept\hbox{${< \atop\hbox{\raise4truept\hbox{$\sim$}}}$}\,}
\newcommand{\gsim}{\, \lower2truept\hbox{${> \atop\hbox{\raise4truept\hbox{$\sim$}}}$}\,}

\newcommand{\hr}{^h}
\newcommand{\mn}{^m}

\begin{document}

\title{ALMA Observations of Warm Dense Gas in NGC~1614 ---
 Breaking of Star Formation Law in the Central kpc}
\thanks{The National Radio Astronomy Observatory is a facility of the National Science Foundation operated under cooperative agreement by Associated Universities, Inc.}
\author{
C.~K.~Xu\altaffilmark{1}, 
C.~Cao\altaffilmark{2,3,1}, 
N.~Lu\altaffilmark{1},
Y.~Gao\altaffilmark{4},
T.~Diaz-Santos\altaffilmark{1},
R.~Herrero-Illana\altaffilmark{13},
R.~Meijerink\altaffilmark{5},
G.~Privon\altaffilmark{6},
Y.-H.~Zhao\altaffilmark{1,4},
A.~S.~Evans\altaffilmark{6,7},
S.~K\"{o}nig\altaffilmark{16},
J.~M.~Mazzarella\altaffilmark{1},
S.~Aalto\altaffilmark{14},
P.~Appleton\altaffilmark{1},
L.~Armus\altaffilmark{1},
V.~Charmandaris\altaffilmark{10,11,12},
J.~Chu\altaffilmark{8},
S.~Haan\altaffilmark{9},
H.~Inami\altaffilmark{15},
E.~J.~Murphy\altaffilmark{1},
D.~B.~Sanders\altaffilmark{8},
B.~Schulz\altaffilmark{1},
P.~van~der~Werf\altaffilmark{5}
}
\altaffiltext{1}{Infrared Processing and Analysis Center, MS 100-22, California Institute of Technology, Pasadena, CA 91125}
\altaffiltext{2}{School of Space Science and Physics, Shandong University at Weihai, Weihai, Shandong 264209, China}
\altaffiltext{3}{Shandong Provincial Key Laboratory of Optical Astronomy and Solar-Terrestrial Environment, Weihai, Shandong 264209, China}
\altaffiltext{4}{Purple Mountain Observatory, Chinese Academy of Sciences, 2 West Beijing Road, Nanjing 210008, China}
\altaffiltext{5}{Leiden Observatory, Leiden University, P.O. Box 9513, NL-2300 RA Leiden, Netherlands}
\altaffiltext{6}{Department of Astronomy, University of Virginia, P.O. Box 400325, Charlottesville, VA 22904, USA}
\altaffiltext{7}{National Radio Astronomy Observatory, Charlottesville, VA 22904, USA}
\altaffiltext{8}{Institute for Astronomy, University of Hawaii, 2680 Woodlawn Dr., Honolulu HI 96816, USA}
\altaffiltext{9}{CSIRO Astronomy and Space Science, ATNF, PO Box 76, Epping 1710, Australia}
\altaffiltext{10}{Department of Physics, University of Crete, GR-71003, Heraklion, Greece}
\altaffiltext{11}{Institute for Astronomy, Astrophysics, Space Applications \& Remote Sensing, National Observatory of Athens, GR-15236, Penteli, Greece}
\altaffiltext{12}{Chercheur Associ\'e, Observatoire de Paris, F-75014,  Paris, France}
\altaffiltext{13}{Instituto de Astrof\'isica de Andaluc\'ia - CSIC, Glorieta de la Astronom\'ia, s/n, 18008, Granada, Spain}
\altaffiltext{14}{Department of Earth and Space Sciences,
Onsala Observatory, Chalmers University of Technology, SE-439 92 Onsala,
Sweden}
\altaffiltext{15}{National Optical Astronomy Observatory,
	950 N. Cherry Ave., Tucson, AZ 85719, USA}
\altaffiltext{16}{Institut de Radioastronomie Millim\'etrique (IRAM), 
300 rue de la Piscine, Domaine Universitaire, F-38406 Saint
          Martin d'H\`eres, France}
\received{Sept.~11, 2014}
\accepted{Nov.~4, 2014}

\begin{abstract}
We present ALMA Cycle-0 observations of the CO~(6-5) line emission and
of the 435~$\mu m$ dust continuum emission in the central kpc of
NGC~1614, a local luminous infrared galaxy (LIRG) at a distance of 67.8
Mpc ($\rm 1\arcsec = 329\; pc$).  The CO emission is well resolved by
the ALMA beam ($\rm 0\farcs26\times 0\farcs20$) into a circum-nuclear
ring, with an integrated flux of $\rm f_{CO~(6-5)} = 898\; (\pm 153)
\; Jy\; km\; s^{-1}$, which is $\rm 63(\pm 12) \%$ of the total
CO~(6-5) flux measured by Herschel. The molecular ring,
located between $\rm 100\; pc < r < 350\; pc$
from the nucleus, looks clumpy and
includes seven unresolved (or marginally resolved) knots with median
velocity dispersion of $\rm \sim 40\; km\; s^{-1}$.  These
knots are associated with strong star formation regions with $\rm
\Sigma_{SFR}\sim 100\; M_\sun\; yr^{-1}\; kpc^{-2}$ and $\rm
\Sigma_{Gas}\sim 10^4\; M_\sun\; pc^{-2}$. The non-detections of the
nucleus in both the CO~(6-5) line emission 
and the 435 $\mu m$ continuum rule out,
with relatively high confidence, a Compton-thick AGN in NGC~1614.
Comparisons with radio continuum emission show a strong deviation from
an expected local correlation
between $\rm \Sigma_{Gas}$ and $\rm \Sigma_{SFR}$, indicating a breakdown
of the Kennicutt-Schmidt law on the linear scale of $\sim$100~pc.  
\end{abstract}


\vskip3truecm

\section{Introduction}
Luminous infrared galaxies (LIRGs) with 
$\rm L_{IR} [8 \hbox{--} 1000\mu m] > 10^{11.5} \; L_\sun$, including
ultra-luminous infrared galaxies (ULIRGs: $\rm L_{IR} > 10^{12}\;
L_\sun$) are mostly advanced mergers \citep{Sanders1988b, Sanders1996, 
Scoville2000, Veilleux2002}. They harbor
extreme starbursts (star formation rate (SFR) $\rm \gsim 50\; M_\sun\;
yr^{-1}$) and sometimes strong active galactic nuclei (AGN), and
are among the most luminous objects in the local Universe
\citep{Sanders1988b, Genzel1998, Surace1998, Veilleux1999, Scoville2000, 
Veilleux2009}.
Observations and theoretical simulations have shown that
mergers can transform spirals to
ellipticals \citep{Toomre1977, Schweizer1982, Barnes1990,
Genzel2001, Veilleux2002, Dasyra2006}.
Gas outflows ubiquitously found in (U)LIRGs  
\citep{Armus1990, 
Heckman2000, Walter2002, Rupke2005,  Sakamoto2009, Fischer2010,
Feruglio2010, Sturm2011, Aalto2012b, Veilleux2013, Cicone2014}
may play an important role in 
quenching the star formation that leads to the
formation of red sequence galaxies \citep{Bell2007, Faber2007, Hopkins2008a,
Hopkins2013b}.

\begin{deluxetable*}{cccccccc}
\tabletypesize{\normalsize}
\setlength{\tabcolsep}{0.03in} 
\tablecaption{ALMA Observations \label{tbl:obs}}
\tablehead{
 {SB}  & Date & {Time (UTC)}  & {Config}  & {$\rm N_{ant}$} & {$\rm l_{max}$} & {$\rm t_{int}$} & {$\rm T_{sys}$}\\
       & (yyyy/mm/dd) &  &   &  & (m) & {(min)} & {(K)}\\
{(1)} & {(2)} & {(3)} & {(4)} & {(5)} & {(6)} & {(7)}  & {(8)}
}
\startdata
X49990a\_X505 & 2012/08/13 & 11:31:46 -- 12:52:33 & E\&C & 23 & 402&24.7 &537 \\
X4b58a4\_X1ee & 2012/08/28 & 08:58:50 -- 10:23:37 & E\&C & 27 & 402&24.7 &756 
\enddata
\tablecomments{Column (1) -- schedule-block number; (2) \& (3) -- observation
date and time; (4) -- configuration; (5) -- number of antennae;
(6) -- maximum baseline length; (7) -- on-target integration time;
(8) -- median $\rm T_{sys}$.
}
\end{deluxetable*}

Extensive surveys of CO rotation lines in low J transitions such as
CO~(1-0) at 2.6 mm and CO~(2-1) at 1.3 mm have found very large concentrations
of molecular gas (up to a few times $\rm 10^{10}\; M_\sun$) in the
central kpc of (U)LIRGs \citep{Solomon1988, Scoville1989,
  Sanders1991, Solomon1997, Downes1998, Bryant1999, Gao2001a,
  Evans2002}. This gas, funneled into the nuclear region by the gravitational
torque during a merger \citep{Barnes1996, Hopkins2009a}, provides fuel for
the nuclear starburst and/or AGN. However, due to the heavy dust
extinction for the UV/optical/NIR observations and the lack of high
angular resolution FIR/sub-mm/mm observations, it is
still not very clear how the different constituents (i.e., gas, dust,
stars, and black holes) in (U)LIRG nuclei interplay with each
other. Some studies \citep{Scoville1997, Downes1998, Bryant1999,
Gao2001b} suggest that much of the low J CO luminosities may be due to
the emission of diffuse gas not closely related to the active star
formation regions.  Indeed, single dish and interferometry mm and
submm observations have found that the intensities and spatial
distributions of star formation in (U)LIRGs correlate significantly
stronger with those of higher J CO lines (with upper level J $\geq
3$), which probe warmer and denser gas than low J lines
\citep{Yao2003, Iono2004, WangJ2004, Wilson2008, Iono2009,
  Sakamoto2008, Tsai2012, Sakamoto2013, Xu2014}. 
This is consistent with results
of observations of other dense molecular gas indicators such as HCN
lines \citep{Solomon1992, Gao2004a, Narayanan2008, Gracia-Carpio2008,
Garcia-Burillo2012}.  The multi-J CO observations of
\citet{Papadopoulos2012a} indicate that for many (U)LIRGs the global CO
spectral line energy distribution (SLED) is dominated by a very warm
($\rm T \sim 100 K$) and dense ($\rm n \geq 10^4\; cm^{-3}$) gas
phase. \citet{Lu2014} found a strong and linear correlation between
the mid-J (with upper level J between 5 and 10) luminosity and the
$\rm L_{IR}$ in a Herschel SPIRE Fourier Transform Spectrometer (FTS)
survey of a large (U)LIRG sample.

In order to study the warm dense gas in nuclear regions of (U)LIRGs,
we observed the CO~(6-5) line emission
(rest-frame frequency = 691.473 GHz) and associated dust continuum
emission in two nearby examples, NGC~34 and NGC~1614, using the Band 9
receivers of the Atacama Large Millimeter Array (ALMA;
\citealt{Wootten2009}).
Both NGC~34 and NGC~1614 were chosen for these early ALMA observations,
among the complete sample of 202 LIRGs of the Great
Observatories All-sky LIRG Survey (GOALS; \citealt{Armus2009}),
because of their close proximity ($\rm D < 85\; Mpc$) and bright
CO~(6-5) line fluxes ($\rm f_{CO~(6-5)} \gsim 1000\; Jy\; km\; s^{-1}$)
observed in the Herschel SPIRE FTS 
survey of GOALS galaxies (angular resolution: $\sim 30\arcsec$;
\citealt{vanderWerf2010, Lu2014}).
 This enables high signal-to-noise-ratio ALMA
observations of warm gas structures with 
linear resolutions of  $\rm \lsim 100\; pc$ for the given  
angular resolutions of $\rm \sim 0\farcs25$.
Further, both LIRGs have declination angles close to the latitude of
the ALMA site, therefore the Band 9 observations are affected by minimal
atmospheric absorption when being carried out near transit.

In this paper, we present ALMA Cycle-0 observations 
of the CO~(6-5) line 
emission and the 435$\mu m$ dust continuum emission in the
central kpc of NGC~1614 (also known as Mrk~617 and Arp~186). 
This LIRG has an infrared luminosity of
$\rm L_{IR} = 10^{11.65}\; L_\sun$ \citep{Armus2009} at a distance of
67.8 Mpc ($\rm 1\arcsec = 329\; pc$). 
Most of the current star formation activity is
in a circum-nuclear starburst ring \citep{Neff1990,
Alonso-Herrero2001, Diaz-Santos2008, Olsson2010}, presumably
triggered by a minor merger with a mass ratio of $\gtrsim 4:1$  
\citep{Neff1990, Vaisanen2012}. The nucleus itself
may harbor a much weaker and older starburst \citep{Alonso-Herrero2001}
and a Compton-thick AGN (\citealt{Risaliti2000}, but see 
\citealt{Olsson2010, Vaisanen2012}). 
The observations and data reductions are described in Section 2;
the results are presented in Section 3;
Section 4 and Section 5 are devoted
to a discussion and the summary, respectively.
All velocities in this paper are in the radio LSR convention.
Throughout this paper, we adopt the $\rm \Lambda$-cosmology with
$\rm \Omega_m = 0.3$ and $\Omega_\Lambda = 0.7$, and  
$\rm H_0 = 70\; km\; s^{-1}\; Mpc^{-1}$.

\begin{figure*}[!htb]
\plottwo{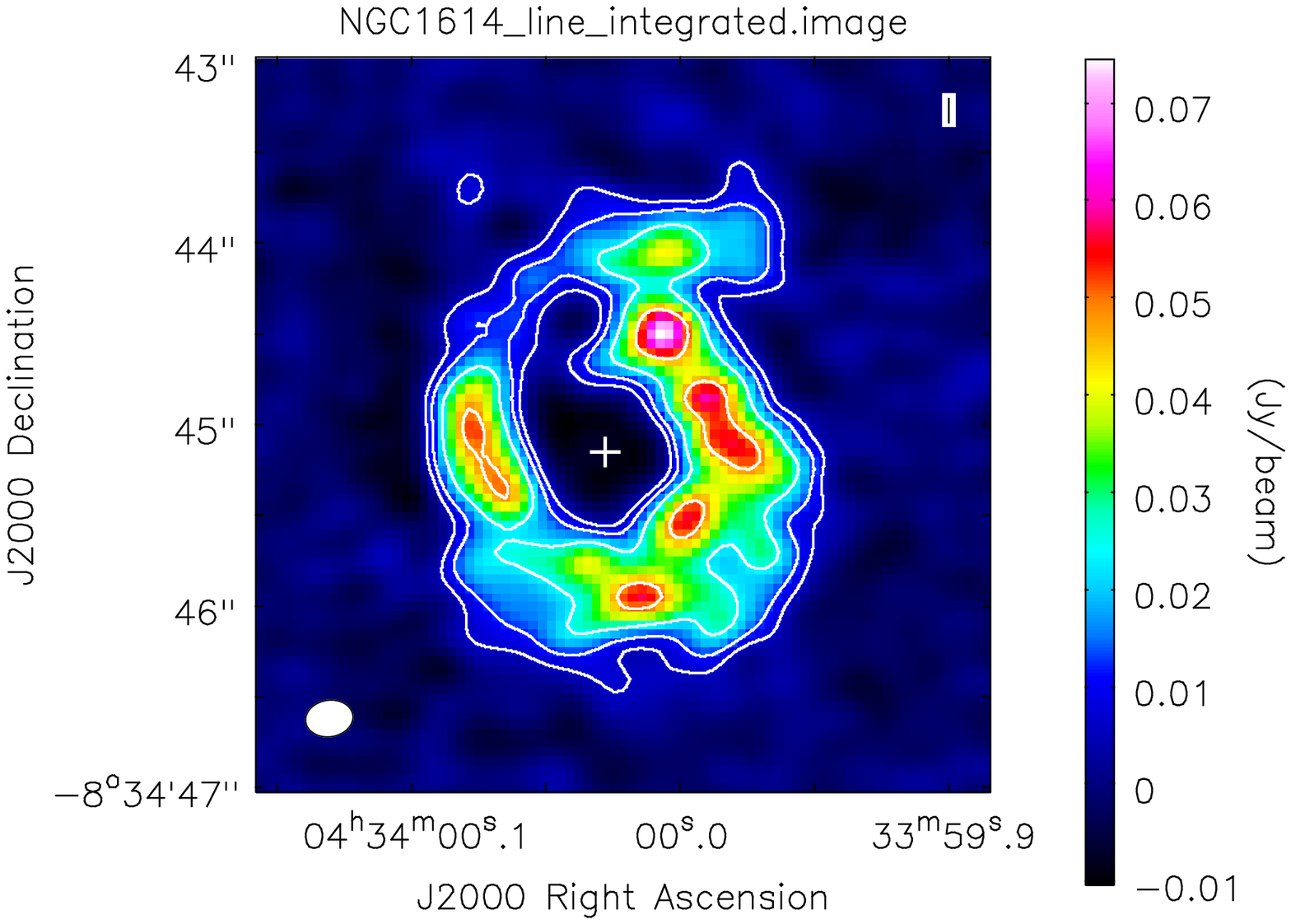}{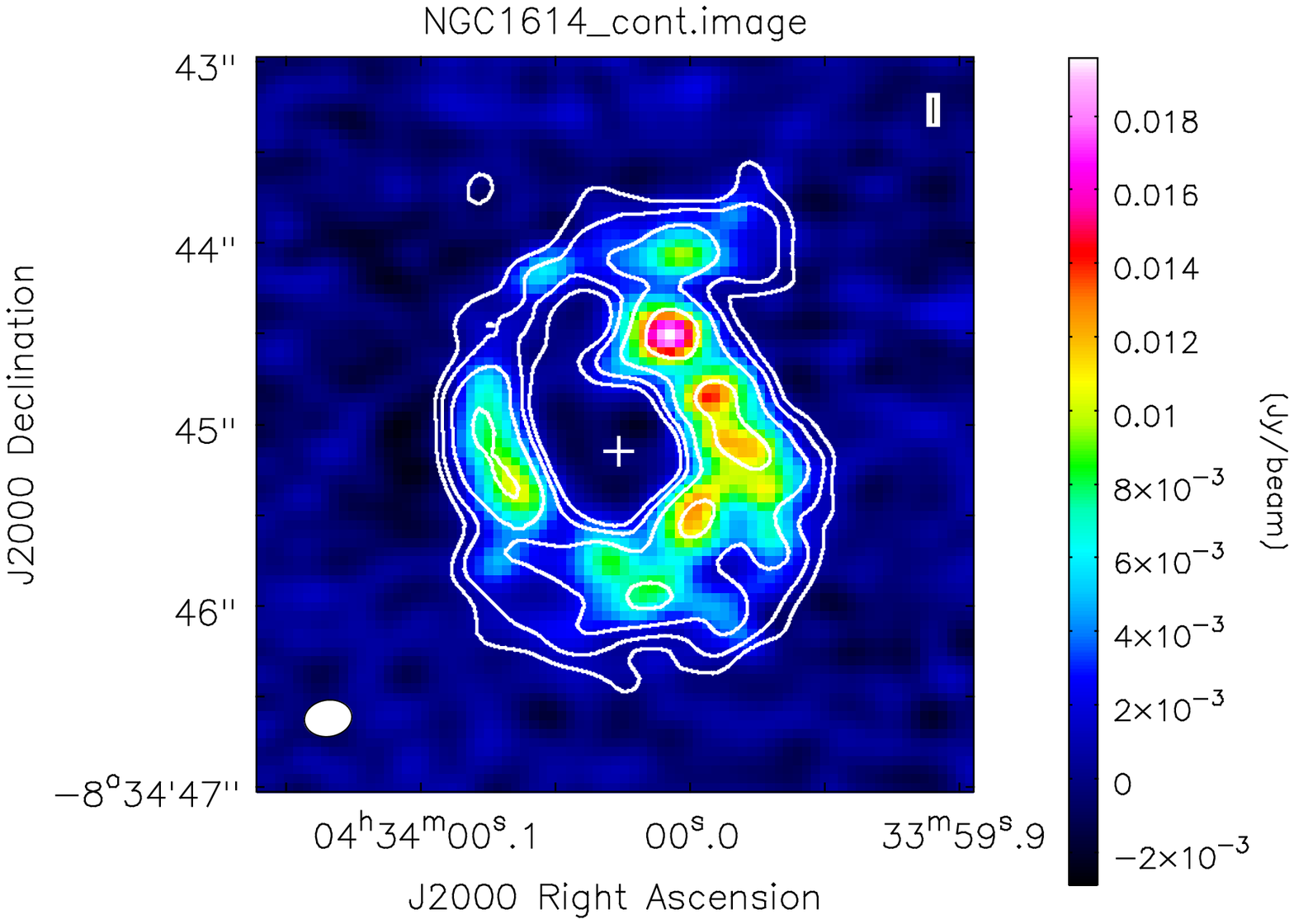}
\plottwo{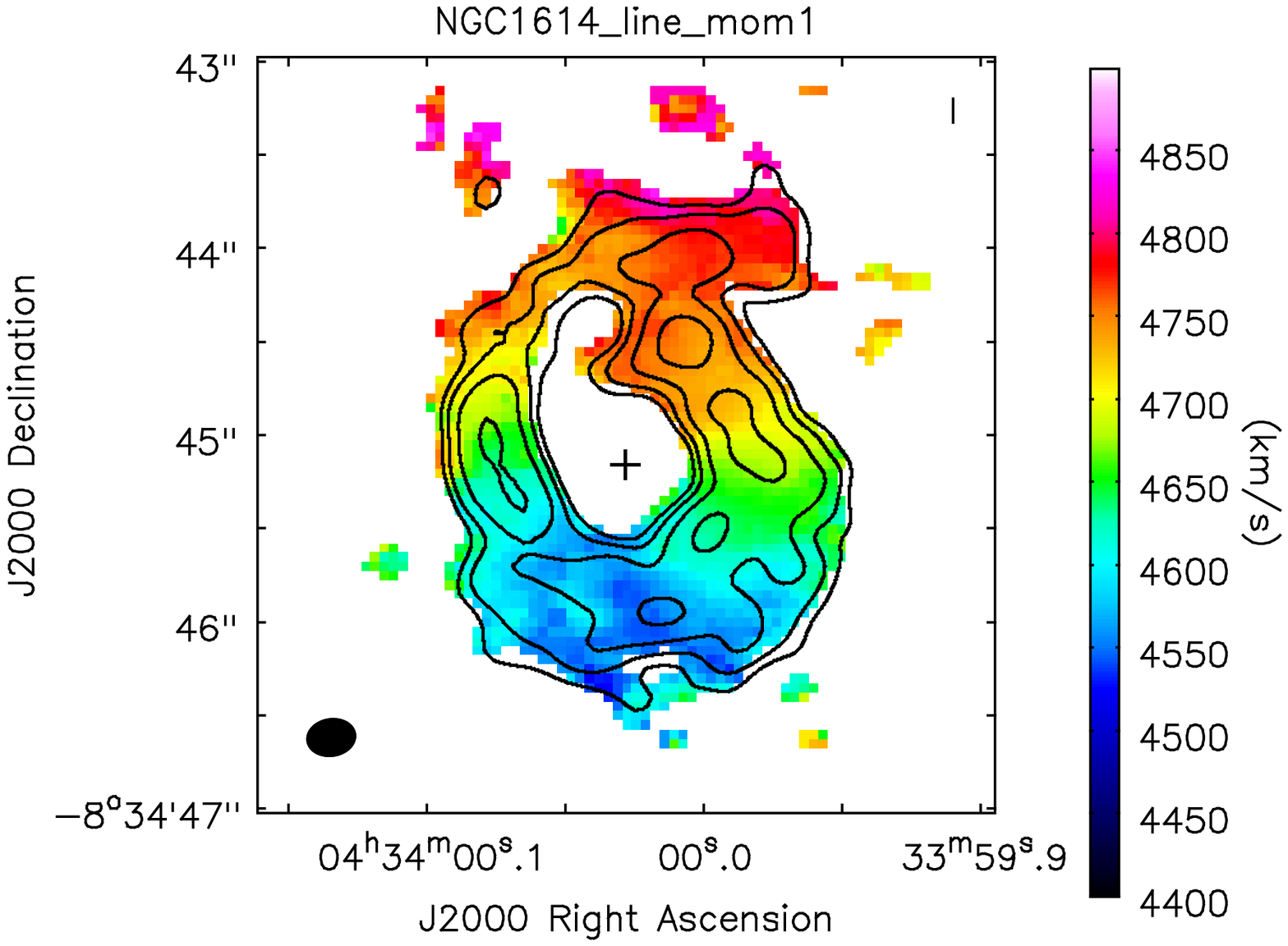}{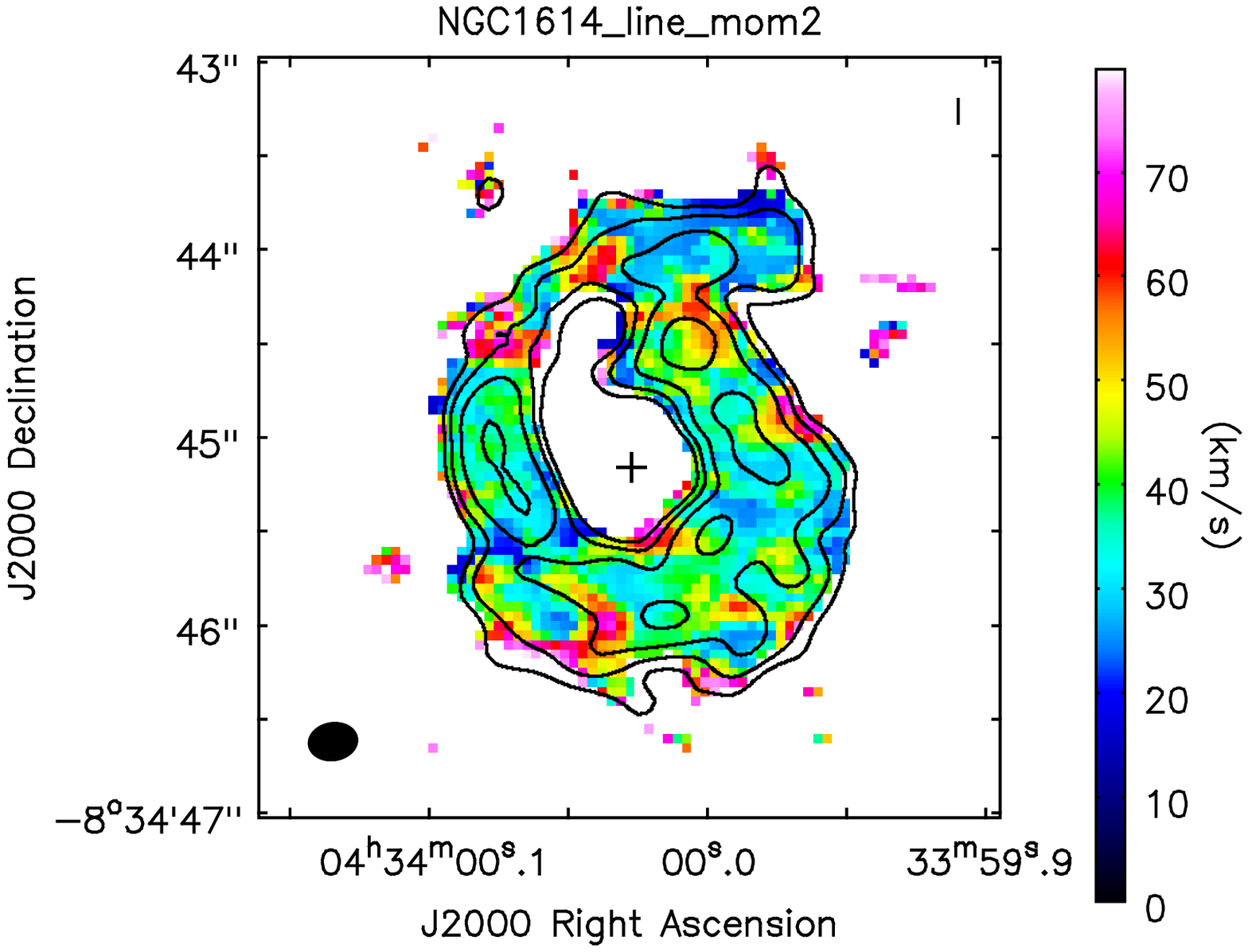}
\caption{
{\it Upper-left:} Image and contours of
the integrated CO~(6-5) line emission. The contour levels are 
[1, 2, 4, 8]$\rm \times 2.7\; Jy\; km\; s^{-1}\; beam^{-1}$.
{\it Upper-right:} Image of the continuum  
overlaid by contours of the integrated CO~(6-5) line emission.
{\it Lower-left:} The first moment map overlaid by contours of 
the integrated CO~(6-5) line emission.
{\it Lower-right:} The second moment map overlaid by contours of 
the integrated CO~(6-5) line emission.
The white (black) ellipse at the bottom left
of each panel shows the synthesized beam size (FWHM =
$\rm 0\farcs26\times 0\farcs20$, $\rm P.A.=280^\circ$). 
All figures have the same size of $\rm 4\arcsec \times 4\arcsec$, and
$\rm 1\arcsec  = 329\; pc$. The cross in the center of each map marks
the position of the radio nucleus in the 5~GHz (MERLIN) map \citep{Olsson2010}.
}
\label{fig:images}
\end{figure*}

\begin{figure*}[!htb]
\plotone{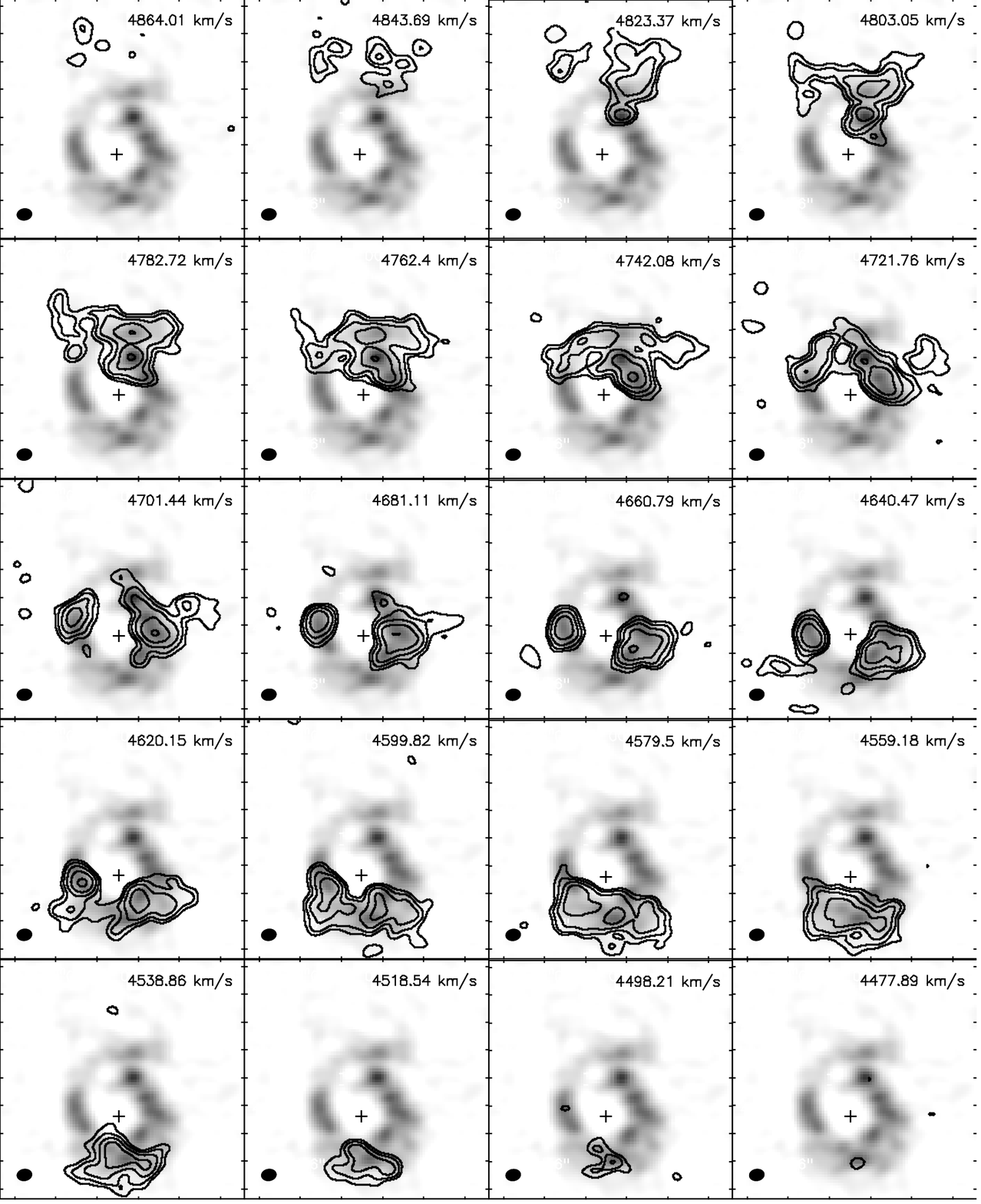}
\caption{CO~(6-5) line emission contours of the channel maps 
(the velocity channel width = $\rm 20.4\; km\; s^{-1}$), and
overlaid on the integrated emission map. The contour levels are
$\rm 21\; mJy\; beam^{-1} \times$ [1, 2, 4, 8, 16]. All maps have the same
size of $\rm 4\arcsec\times 4\arcsec$. In each panel, the central velocity
of the channel is given. (The system velocity of NGC~1614 is $\rm 4723\; km\; s^{-1}$.)}
\label{fig:channel}
\end{figure*}

\section{Observations}
We observed the central region of NGC~1614 in
CO~(6-5) line emission and 435~$\mu m$ dust continuum
emission using the
Band 9 receivers of ALMA in the time division mode 
(TDM; velocity resolution: 6.8
km~sec$^{-1}$). The four basebands (i.e. ``Spectral Windows'',
hereafter SPWs) were centered at the sky frequencies of 680.539,
682.308, 676.826 and 678.764 GHz, respectively, each with a bandwidth
of 2 GHz. Observations were carried out in the 
extended \& compact (E\&C) configuration using up to 27 antennae 
(Table~\ref{tbl:obs}).
The total on-target integration time was 50.4 minutes. 
During the observations, phase and gain variations
were monitored using QSO 0423-013. Observations of
the minor planet Ceres were made for the flux calibration. 
The error in the flux calibration was estimated to be 17\%.

The final data reduction was done using CASA 4.1.0. Images were cleaned using
Briggs weighting. Both phase and amplitude self-calibrations have 
been carried out. The primary beam is $\sim 8\arcsec$. However, 
emission features larger than
$\sim 3\arcsec$ are poorly sampled because of limited uv-coverage for
short baselines. 
Two data sets were generated from the
observations. In the first data set, the CO~(6-5) line data cube was
generated using data in SPW-0 (sky-freq = 680.539$\pm 1$ GHz),
which encompass the CO~(6-5) line emission at the system velocity
($\rm 4723\; km\; s^{-1}$) with an effective bandpass of
$\rm \sim 800\; km\; s^{-1}$.  The continuum was estimated using data in
the other 3 SPWs. In the second data set, the CO~(6-5) line data
cube was generated using three SPWs: SPW-0, SPW-1 and SPW-3, 
respectively. The bandwidth of
this CO~(6-5) line data cube is $\rm \sim 2400\; km\; s^{-1}$.
For the second data
set, the continuum was estimated using data in SPW-2 (sky-freq =
676.826$\pm 1$ GHz). The first data set 
has better continuum estimation and subtraction,
and was therefore used for most of the analysis; the second data set was used to
search for evidence of molecular outflows or nuclear $\rm H^{13}CN$~(8-7) 
line emission. The continuum subtraction
was carried out using the CASA task {\it UVcontsub}. For the first data set,
the channel maps of the CO~(6-5) line emission have 1-$\sigma$
rms noise of $\rm 8.0\; mJy\; beam^{-1}$ per $\rm 6.8\; km\; s^{-1}$,
and for the continuum it is $\rm 0.6\; mJy\; beam^{-1}$.
The CO~(6-5) line emission map, 
integrated over the LSR velocity range
between $\rm v = 4447.4$ and $\rm 4894.5\; km\; s^{-1}$
($\rm \delta v = 447.1\; km\; s^{-1}$), has 1-$\sigma$
rms noise of $\rm 0.90\;  Jy\; beam^{-1} \; km\; s^{-1}$.
The synthesized beams
of these maps are nearly identical, having FWHMs of $\rm
0\farcs26\times 0\farcs20$, corresponding to physical scales of
$\rm 86\; pc \times 66\; pc$, and a P.A. of $280^\circ$.  The absolute
pointing accuracy of these ALMA observations is on the order of
$0\farcs1$.

\begin{figure}[!htb]
\epsscale{1.1}
\plotone{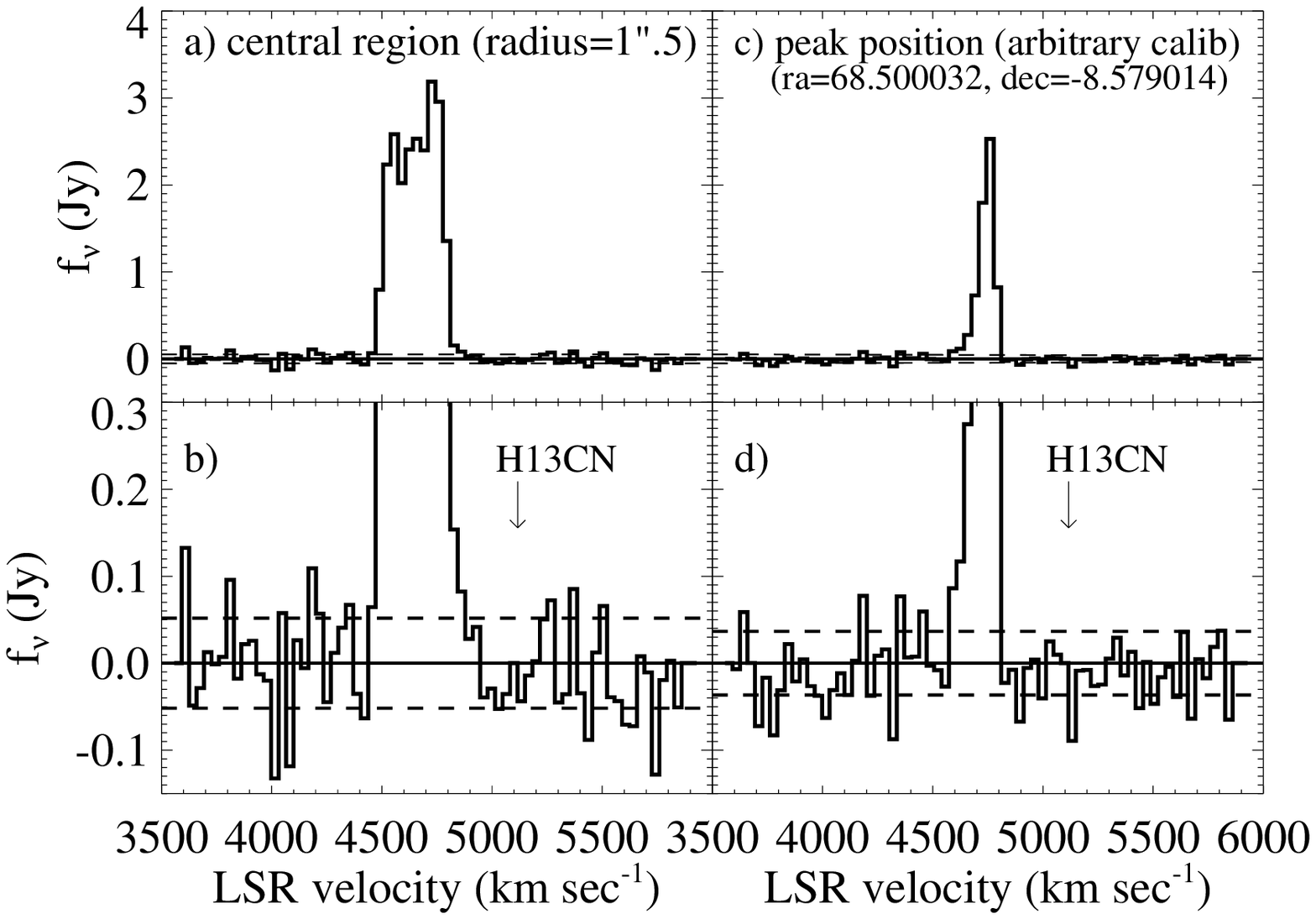}
\caption{
{\bf Panel a:} Spectrum of the CO~(6-5) line emission
in the velocity domain, measured in the channel maps
with an aperture of $\rm radius=1\farcs5$. The dashed lines mark the
1-$\sigma$ noise boundaries.  
{\bf Panel b:} Zoom-in of the bottom part of {\it Panel a}. 
Again the dashed lines mark the 1-$\sigma$ noise boundaries. The arrow
marks the expected 
location of the $\rm H^{13}CN$~(8-7) line at the systemic velocity
of $\rm v = 4723\; km\; s^{-1}$.
{\bf Panel c:} Spectrum at the peak position 
($\rm ra=04:34:00.006$, $\rm dec=-08:34:44.47$)
of the integrated CO~(6-5) line emission map. In order to show it more
clearly, the flux is scaled up arbitrarily.
{\bf Panel d:} Zoom-in of the bottom part of {\it Panel c}. 
}
\label{fig:spectra}
\end{figure}

\section{Results}
\subsection{The CO~(6-5) Line Emission}\label{sect:CO}
In Figure~\ref{fig:images} we present images of the integrated 
CO~(6-5) line emission, 
the continuum at 435~$\mu m$, the first moment map, and the
second moment map. All images are overlaid by the same contours
of the integrated CO~(6-5) line emission at levels of
[1, 2, 4, 8]~$\rm \times 2.7\; Jy\; beam^{-1} \; km\; s^{-1}$.
The line and the continuum emissions correlate closely with each other,
both showing a ring configuration without a detectable nucleus. 
The ring has a diameter of $\sim 2\arcsec$ ($\rm \sim 650\;
pc$). In both the line and the continuum maps, the ring 
looks clumpy and can be decomposed into several knots.
The first moment map shows a clear velocity gradient along the 
south-north direction, consistent with an inclined rotating ring.
According to \citet{Olsson2010}, the inner disk of NGC~1614 has 
an inclination angle of $\rm 51^\circ$.
In the second moment map, the velocity
dispersion in most regions in the ring is rather constant at the level of 
$\rm \delta v \sim 40\;  km\; s^{-1}$, though in some inter-knot regions
it can be as low as $\rm \delta v \sim 20\;  km\; s^{-1}$.
From the first moment map, the velocity gradient
due to the rotation can be estimated to be 
$\rm dV/dr \sim 0.3 \; km\; s^{-1}\; pc^{-1}$,
corresponding to a line widening of $\sim 20\; km\; s^{-1}$ within 
individual beams (linear size: $\sim 80\; pc$). This is 
consistent with the lowest velocity dispersion seen in the
second moment map.

The channel maps ($\rm \delta v = 20.4\; km\; s^{-1}$) are shown
in Figure~\ref{fig:channel}, overlaid on the image of the
integrated line emission. They provide more details about the rotating
ring. First of all, given the relatively narrow local velocity dispersions
(see the second moment map in Figure~\ref{fig:images}), the channel maps 
dissect the ring spatially. It appears that the spatial width of ring segments
in individual channel maps are generally broader than that in the
integrated emission map. This is because, by co-adding all channel maps,
the integrated emission map is affected more severely 
by the (negative) sidelobes of different segments of the ring.
This is a significant effect because some ring segments are separated
by $\rm \sim 3\arcsec$, the angular scale limit of our 
interferometer observations. Indeed the total flux of the CO~(6-5) line
emission,  $\rm f_{CO~(6-5)} = 898\; (\pm 153) \; Jy\; km\; s^{-1}$
(the error being dominated by the calibration uncertainty) which is
derived from the sum of the aperture photometry of individual channels 
(centered on the emission features for each given channel), is 31\% higher
than that measured on the integrated CO~(6-5) line emission map.
Comparison with the Herschel measurement of the
  integrated CO~(6-5) line emission of NGC~1614 ($\rm 1423\pm 126\; Jy\;
  km\; s^{-1}$ within a beam of $\sim 30\arcsec$; 
\citealt{vanderWerf2010, Lu2014}) yields an
  interferometer-to-single-dish flux ratio of $0.63\pm 0.12$.
This suggests that most warm dense gas in NGC~1614 is concentrated
in the circum-nuclear ring. 

Figure~\ref{fig:spectra} shows plots of
the velocity distributions of the central region
($\rm radius= 1\farcs5 \simeq 500\; pc$) and of the peak position 
($\rm RA=04{\hr}34{\mn}00{\fs}006$, $\rm Dec=-08{\degr}34{\arcmin}44{\farcs}47$)
of the integrated CO~(6-5) line emission map. 
In order to reduce the noise, we used relatively broad bins
of $\rm \delta v = 34\; km\; s^{-1}$.  
The velocity distribution has a FWHM of 
$\rm 272\; km\; s^{-1}$. Its shape is rather irregular and
spiky, reflecting the clumpiness of the rotating ring and the
narrow velocity dispersions of individual clumps 
(Figure~\ref{fig:images}). No evidence for any outflow/inflow of
$\rm |\delta v| < 1200\; km\; s^{-1}$, nor  
any detection of the $\rm H^{13}CN$~(8-7) line 
(rest-frame frequency = 690.552 GHz), can be found in the spectrum.
In the velocity distribution of the peak position, which is
in the north-western quadrant of the ring (Figure~\ref{fig:images}),
we also found no evidence of outflow/inflow or of the $\rm H^{13}CN$~(8-7) 
line emission.

\begin{figure}[!htb]
{\hspace{-0.8truecm}\epsfig{figure=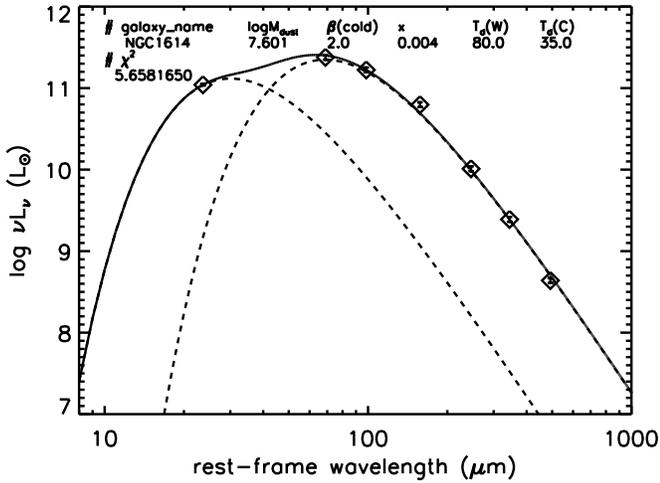, angle=90, scale=0.4}}
\caption{SED fitting of the dust emission in NGC~1614 
by a 2-graybody model.}
\label{fig:dust_sed}
\end{figure}


\begin{figure}[!htb]
\epsscale{1.15}
\plottwo{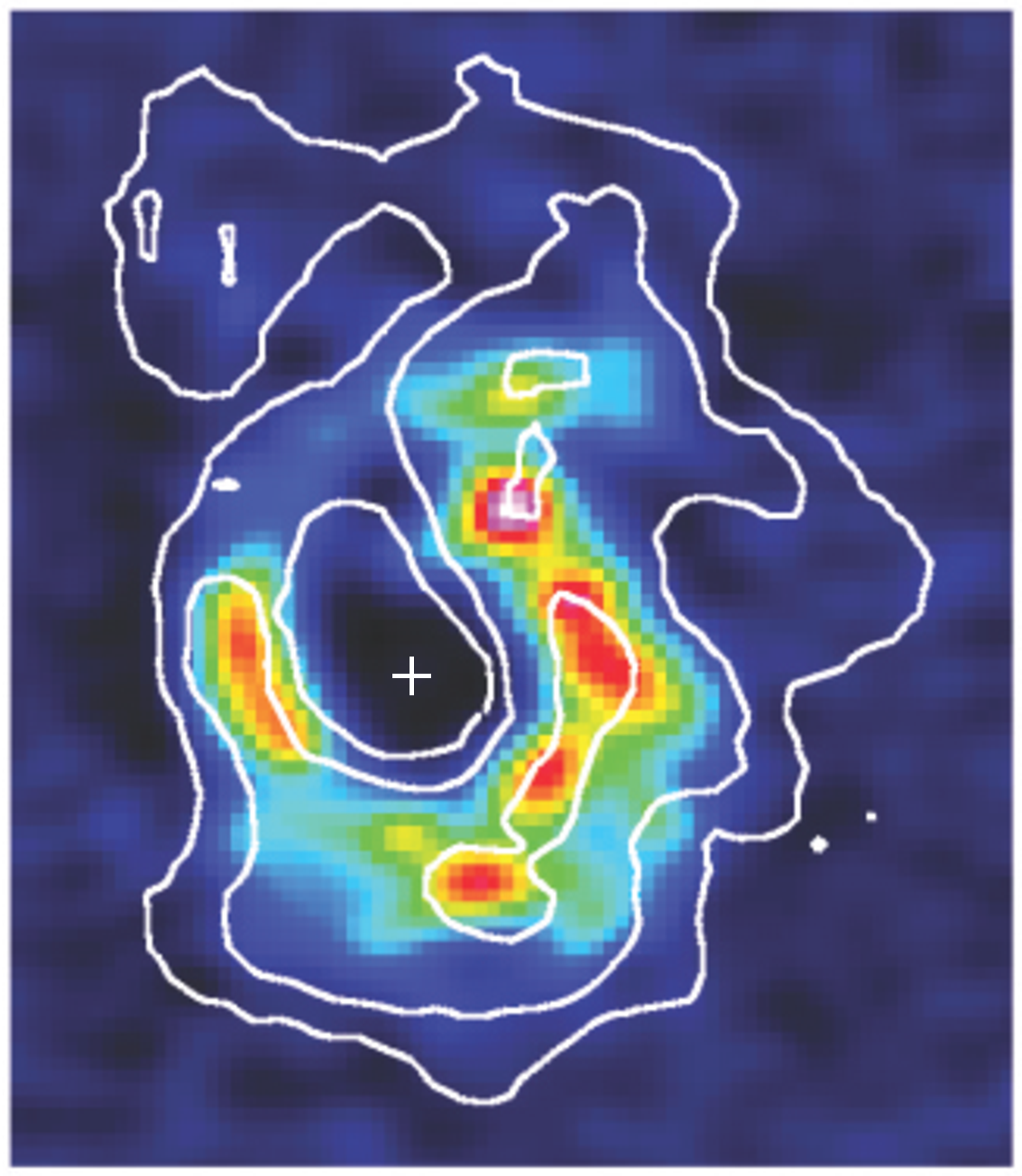}{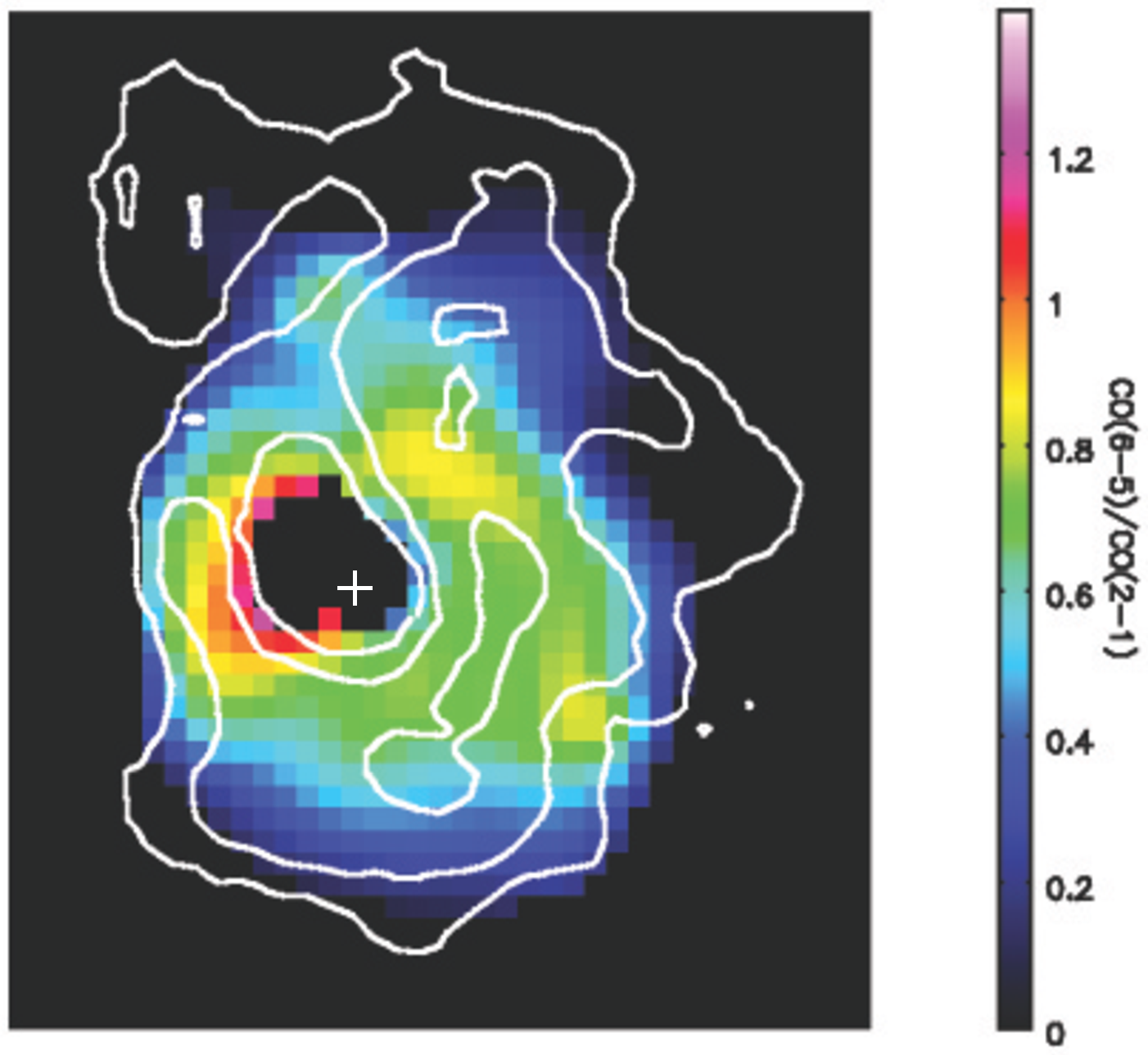}
	\caption{
{\bf Left:} Comparison between 
integrated CO~(6-5) line emission map (resolution: $0\farcs26\times 0\farcs20$)
and
integrated CO~(2-1) line emission contours 
(resolution: $0\farcs50\times 0\farcs44$; \citealt{Konig2013}).
{\bf right:} Contours of 
integrated CO~(2-1) line emission overlaid on
image of the ratio between
integrated CO~(6-5) line emission and
integrated CO~(2-1) line emission, with 
both being smoothed to a common beam
(convolution of two original beams). Signals in the both maps are in the
same units of $\rm K\; km\; s^{-1}$.
}
\label{fig:compCO21}
\end{figure}

\begin{figure*}[!htb]
\epsscale{1.1}
\plotone{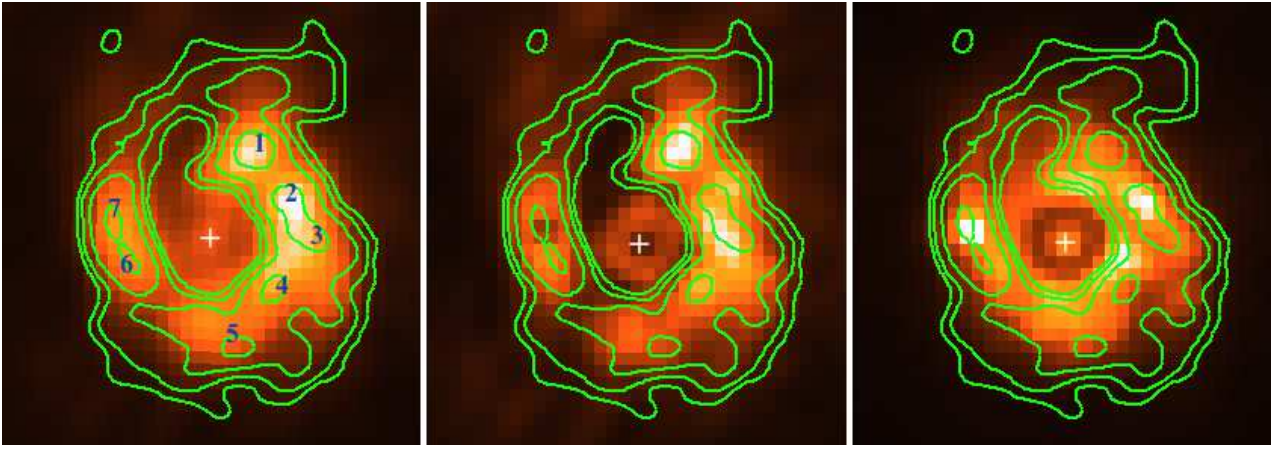}
\caption{Comparison between
contours of the integrated CO~(6-5) line emission and 
images of the total 8.4 GHz radio continuum (left),
the nonthermal radio component (middle), and the thermal radio 
component (right). Positions of CO~(6-5) knots listed in Table~\ref{tbl:knots}
are marked by corresponding numbers in the left panel.
}
\label{fig:sfr}
\end{figure*}

\subsection{The 435~$\mu m$ Continuum Emission}\label{sec:dust}

The flux density of
the 435~$\mu m$ continuum measured by ALMA is $\rm f_{435\mu m}=269\pm 46\; mJy$.
The continuum correlates spatially with the CO~(6-5)
emission in the central kpc of NGC~1614 (Figure~\ref{fig:images}). 
This
suggests that dust heating and gas heating in the warm dense gas cores
are strongly coupled, a conclusion also reached by \citet{Lu2014}
in a Herschel FTS study of the CO SLED of LIRGs. 
NGC~1614 was observed by Herschel-SPIRE \citep{Griffin2010} both
in the photometry mode (Chu et al., in preparation) and in
the FTS mode \citep{vanderWerf2010, Lu2014}, 
with beams of $\rm \sim 30\arcsec$. Because the
error of the continuum measured in the FTS mode is large ($\rm \sim 1\; Jy$),
we estimated the total flux of the 435~$\mu m$ continuum of NGC~1614
using SPIRE photometer fluxes $\rm f_{350\mu m, SPIRE}=1916\pm 134\; mJy$
and $\rm f_{500\mu m, SPIRE}=487\pm 34\; mJy$. 
Assuming a power-law spectrum for the dust continuum
(i.e. $\rm \log f_\nu$ depending on $\rm \log \nu$ linearly), we carried out
linear interpolation in the logarithmic domain of the flux and of the
frequency between 350 and 500~$\mu m$, and found 
$\rm f_{435\mu m, SPIRE}=831\pm 58\; mJy$.
The ratio between $\rm f_{435\mu m}$ and $\rm f_{435\mu m, SPIRE}$ then yields
an interferometer-to-single-dish flux ratio of $\rm 0.32\pm 0.06$. 
This is a factor of $\sim 2$
lower than the interferometer-to-single-dish flux ratio of the line
emission, indicating that the distribution of dust is substantially more
extended than that of the warm dense gas.

The total dust mass
in NGC~1614 can be estimated using the mid- and far-IR fluxes in the
Spitzer/MIPS 24~$\mu m$ band and in the Herschel 70, 100, 160, 250, 350 and
500~$\mu m$ bands. The Herschel data are taken from Chu et al. (in
preparation). 
A least-squares fit to the IR SED by a 2-graybody model, with the
  emissivity spectral index $\rm \beta = 2$ for both components,
  yields a total dust mass of $\rm M_{dust, total} = 10^{7.60\pm
    0.07}\; M_\sun$ with a cold dust temperature of $\rm T_C = 35\pm 2
  K$ (Figure~\ref{fig:dust_sed}). Fits by 2-graybody models with $\rm
  \beta$ as a free parameter or by the model of \citet{DL07} yield
  very similar results.  If dust in the central region 
  has the same $\rm T_C$, then $\rm M_{dust, cent} = f_{435\mu m,
    ALMA}/ f_{435\mu m, SPIRE} \times M_{dust, total}$.  Taking into
  account the uncertainties due to the assumption on the cold dust
  temperature ($\sim 50\%$), the dust mass in the central region
  observed by ALMA is $\rm M_{dust, cent} = 10^{7.11\pm 0.20}\;
  M_\sun$.  NGC~1614 has a metallicity of $\rm 12+log(O/H) = 8.65\pm
  0.10$ \citep{Armus1989, Vacca1992, Engelbracht2008,
    Modjaz2011}. According to \citet{Remy-Ruyer2014}, for galaxies
  with $\rm 12+log(O/H)> 8.5$, the gas-to-dust ratio is 100 with a
  1-$\sigma$ uncertainty of $\sim 0.2$~dex.  Therefore, assuming $\rm
  M_{gas}/M_{dust} = 10^{2.0\pm 0.2}$, the gas mass in the central
  region of NGC~1614 is $\rm M_{gas, cent} = 10^{9.11\pm 0.30}\;
  M_\sun$. This is consistent with the molecular gas mass (which should
  dominate the total gas mass) found in the same region ($\rm M_{gas,
    cent} = 10^{9.30}\; M_\sun$, with a conversion factor of $\rm
  X_{CO} = 3\times 10^{20}\; cm^{-2}(K\; km\; s^{-1})^{-1}$;
  \citealt{Konig2013}). It should be pointed out that both the ALMA 
  continuum observations and the SMA observations of CO~(2-1) by 
  \citet{Konig2013} detected mostly dust and gas emission in dense gas
  structures and missed significantly the diffuse emission, therefore the dust
  and gas mass derived from these observations are lower limits.

  There has been a debate on whether there is an AGN in
  NGC~1614. \citet{Risaliti2000} argued, based on the detection of a
  hard X-ray source and its spectrum, that in the center of NGC~1614
  there is a hidden AGN obscured by Compton-thick gas ($\rm N_{H} >
  1.5\times 10^{24}\; cm^{-2}$; \citealt{Comastri2004}).  However such
  high column density gas in the nucleus, which shall not be affected
  by the missing flux issue, is not detected in either the CO~(6-5)
  map or the dust continuum map.  Using the scaling factor between the
  gas mass and the continuum flux derived above, the non-detection of
  the continuum in the nucleus ($\rm \sigma = 0.6\; mJy\; beam^{-1}$)
  sets a 3-$\sigma$ upper-limit for the gas surface density of $\rm
  N_{H} = 10^{23.1\pm 0.3}\; cm^{-2}$.
Since an AGN cannot hide itself in the emission of
dust that absorbs its UV/optical/NIR radiation, our results can rule
out this possibility with relatively high confidence. Indeed a
Compton-thick torus with a radius of $\rm r=20\; pc$
\citep{Garcia-Burillo2014}, which fills 23\% of the ALMA beam, should
be detectable in the continuum with a signal-to-noise ratio of $\rm 
s/\sigma\geq 7$. The $\rm s/\sigma$ could be even higher 
since the $\rm T_d$ in a torus is likely much 
warmer than the assumed dust temperature of $\rm T_d= 35\; K$.
The high resolution MIR L-band observations of
\citet{Vaisanen2012} also argue against a Compton-thick AGN in
NGC~1614. As pointed out by \citet{Olsson2010} and
\citet{Herrero-Illana2014}, the X-ray source detected by
\citet{Risaliti2000} could be explained by low-mass X-ray binaries.

\section{Discussion}
\subsection{Comparison with Previous CO Observations}\label{sect:co_comp}
There is a rich literature on molecular line observations in
the submm and mm bands for NGC~1614 \citep{Young1986, Solomon1988,
Scoville1989, Sanders1991, Casoli1991, Gao2004b, Albrecht2007, Wilson2008,
Olsson2010, Costagliola2011, Konig2013, Imanishi2013}. 
The single dish CO~(1-0) observations
of \citet{Sanders1991} found a total molecular gas mass of
$\rm 10^{10.12}\; M_\sun$, aasuming a standard conversion factor of
$\rm X_{CO} = 3\times 10^{20}\; cm^{-2}(K\; km\; s^{-1})^{-1}$
and distance of $\rm D=67.8\; Mpc$. This is consistent with the
result of \citet{Casoli1991}, but significantly larger than
those obtained in earlier and less sensitive observations 
\citep{Young1986, Solomon1988}. The OVRO observations
of \citet{Scoville1989}, with a beam of $4\arcsec \times 6\arcsec$, 
allocated $\rm 30\%$ of the total CO~(1-0) emission
within a nuclear region of $\rm radius = 1\; kpc$.
The more recent and higher 
resolution ($2\farcs75 \times 2\farcs40$) observations 
of \citet{Olsson2010} resolved the central CO~(1-0) line emission into
an arc-like feature $\rm \sim 3\; kpc$ in length and $\rm \sim 1.3\; kpc$ in width, but did not resolve the ring. The SMA map of CO~(3-2) 
(beam $=2\farcs6\times 2\farcs1$; \citealt{Wilson2008}) 
and the ALMA maps of HCN/HCO$^+$/HNC~(4-3) 
(beam $=1\farcs5\times 1\farcs3$;  \citealt{Imanishi2013}) also 
did not resolve the ring.

Before our ALMA observations, the best angular resolution for any CO
rotation lines was obtained by \citet{Konig2013} in their SMA
observations of the CO~(2-1) line, with a beam of $\rm 0\farcs50\times
0\farcs44$.  In Figure~\ref{fig:compCO21} we compare their CO~(2-1)
map with our CO~(6-5) map. In the ring the two maps have good
correspondence, though the CO~(6-5) emission looks clumpier, most
likely due to the better angular resolution.  \citet{Konig2013}
noticed a strong asymmetry in the CO~(2-1) distribution between the
eastern and western sides of the ring, and interpret it as a consequence of the
feeding of the ring by the dust lane on the north-west of the ring.
In the CO~(6-5) map, we still see
this asymmetry albeit being less prominent than in CO~(2-1). 
When
smoothed to a common beam, the ratio between the two emissions is
rather constant in most regions of the ring, with a median brightness
temperature ratio of 0.72 (Figure~\ref{fig:compCO21}). The east
quadrant of the ring has the highest brightness temperature
ratio ($\sim 1$). This could be due, at least partially,
to a slight mismatch between the two maps, given the steep gradient in both maps
in this region. On the other hand, this seems to be
consistent with the stronger east-west asymmetry seen in the CO~(2-1)
map than in the CO~(6-5) map. \citet{Konig2013} argued that
the reason of the asymmetry could be the
feeding of the ring by a dust lane on the northwest of the ring.
In this scenario, the east quadrant has higher CO~(6-5)/CO~(2-1) ratio
than the west quadrant because it has less diffuse gas (freshly fed by
the dust lane) compared to the west quadrant.
The nucleus is not detected in either map. 
The CO~(6-5) 3-$\sigma$ upper-limit of 742 $\rm
M_\sun\; pc^{-2}$ ($\rm N_{H} = 10^{22.86}\; cm^{-2}$) for the surface
density of the warm dense gas, derived by
assuming the same relation between $\rm \Sigma_{Gas}$ and 
CO~(6-5) surface brightness in the ring region (Eq~\ref{eq:gas}),
is consistent with the upper-limit set by the 435~$\mu m$ continuum. 
If the conversion factor advocated by \citet{Downes1998} for (U)LIRGs
is used, which is a factor of $\sim 6$ lower than the standard
value adopted in Eq~\ref{eq:gas}, the result is a significantly lower
value for the upper limit of $\rm \Sigma_{Gas}$ in the nucleus.
In the region north of the ring, where significant CO~(2-1) emission is
found, little CO~(6-5) emission is detected and the brightness
temperature ratio is $<0.1$.

\begin{deluxetable*}{cccccccc}[ht]
\tabletypesize{\normalsize}
\setlength{\tabcolsep}{0.05in} 
\tablecaption{CO~(6-5) Knots in Circum-Nuclear Starburst Ring \label{tbl:knots}}
\tablehead{
 (1) & (2) & (3) & (4) & (5) & (6) & (7) & (8) \\
\hline
ID & R.A.    & Dec. & $\rm S_{CO~(6-5)}$ & $\rm S_{435\mu m}$ & $\rm \log (\Sigma_{Gas})$& $\rm \log (\Sigma_{SFR})$ & Notes \\
\hline
  & (J2000) & (J2000) & ($\rm Jy\; km\; s^{-1}\; beam^{-1}$)   & ($\rm mJy\; beam^{-1})$ & ($\rm M_\sun\; pc^{-2})$  & ($\rm M_\sun\; yr^{-1}\; kpc^{-2}$)    & 
}
\startdata
  1 & 04:34:00.006 & -08:34:44.49  & 34.0$\pm 6.1$  & 19.6$\pm 3.5$  & 3.94 &  2.49  & 4  \\
  2 & 04:33:59.991 & -08:34:44.86  & 27.2$\pm 4.9$  & 13.8$\pm 2.5$  & 3.87 &  2.40  & 6  \\
  3 & 04:33:59.981 & -08:34:45.10  & 25.5$\pm 4.6$  & 12.1$\pm 2.2$  & 3.91 &  2.33  & 6  \\
  4 & 04:33:59.998 & -08:34:45.52  & 25.5$\pm 4.6$  & 11.7$\pm 2.1$  & 3.88 &  2.41  & 7  \\
  5 & 04:34:00.015 & -08:34:45.95  & 25.7$\pm 4.6$  &  7.7$\pm 1.4$  & 3.87 &  2.14  & 8  \\
  6 & 04:34:00.069 & -08:34:45.29  & 22.8$\pm 4.1$  & 10.3$\pm 1.9$  & 3.81 &  2.36  & 10 \\
  7 & 04:34:00.077 & -08:34:45.04  & 23.6$\pm 4.2$  &  8.2$\pm 1.5$  & 3.84 &  2.00  & 10
\enddata
\tablecomments{
Column (4) -- CO~(6-5) peak flux;
 (5) -- continuum peak flux;
 (6) -- peak molecular gas surface density, after smoothed to the
        beam of 8.4 GHz observations (0\farcs41$\times$0\farcs26);
 (7) -- peak SFR density, derived using flux of the nonthermal radio at 
        8.4 GHz;
 (8) -- corresponding GMA in \citet{Konig2013}.
}
\end{deluxetable*}
\subsection{Relation Between Warm Dense Gas and Star Formation}
Most star formation in NGC~1614 is occurring in the circum-nuclear
starburst ring. \citet{Soifer2001} found that
72\% of the 12$\mu m$ flux of NGC~1614
measured by IRAS is contained within the 2{\arcsec} beam of their
Keck observations. The comparison between the high resolution 8.4 GHz map of the
central kpc region \citep{Herrero-Illana2014}
and that of a lower resolution map at the same frequency by \citet{Schmitt2006}
indicates that $\sim 67\%$ of the total SFR is contributed by
the starburst ring and the nucleus.
The star formation in the ring has been studied by 
\citet{Alonso-Herrero2001} using a 
HST Pa~$\alpha$ map, \citet{Diaz-Santos2008}
using a Gemini 8$\mu m$ map, \citet{Vaisanen2012} using a 3.3$\mu m$  
polycyclic aromatic hydrocarbon (PAH) map obtained with
UIST (an imager-spectrometer for integral field spectroscopy)
at UKIRT, \citet{Olsson2010} and \citet{Herrero-Illana2014} 
using VLA and Merlin 
radio continuum maps.  While the NIR and MIR observations may
still be affected by the dust obscuration associated with the dense gas
\citep{Imanishi2013}, the
radio continuum is an SFR indicator insensitive to dust obscuration.

In the left panel of Figure~\ref{fig:sfr}, we compare the CO~(6-5) contours 
with the radio continuum at 8.4 GHz 
(beam = $\rm 0\farcs41\times 0\farcs26$, \citealt{Herrero-Illana2014}).
The CO~(6-5) knots in the ring (Table~\ref{tbl:knots}) are marked in the image.
While there is a radio nucleus, 
the CO map has a hole at the ring's center.
\citet{Herrero-Illana2014} argued that the radio nucleus is not an
AGN, which is consistent with our conclusion that there is no
(Compton-thick) AGN in NGC~1614 (Section~\ref{sec:dust}).  In the other
two panels of Figure~\ref{fig:sfr}, the CO~(6-5) emission is compared
to the thermal and nonthermal radio emission components, respectively.
Following \citet{Herrero-Illana2014}, the thermal radio emission is estimated
using a high-resolution, extinction-corrected Pa-${\alpha}$ map
obtained from a set of HST NIR narrow- and broad-band images.
The method involves the comparison
of a Pa-${\alpha}$ equivalent width map as well as an NIR color (F160W/F222M) 
image
to stellar population synthesis models (Starburst99; \citealt{Leitherer1999})
to derive a spatially-resolved dust obscuration map with which to correct
the original Pa-${\alpha}$ image (see also \citealt{Diaz-Santos2008}).
The details of this procedure 
can be found in the Appendix. As a final step, the nonthermal
component is derived by subtracting from the total radio emission the thermal component.
The thermal fraction of the radio continuum at 8.4 GHz is found to
be 51\% in the ring region ($\rm 100 < r < 350$~pc).

\begin{figure}[!htb]
\epsscale{1.25}
\plotone{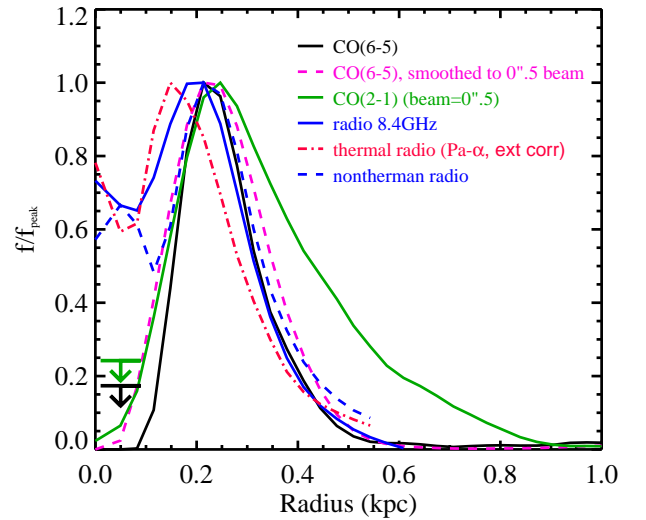}
\caption{Comparison between normalized radial profiles of the CO~(6-5),
total radio continuum at 8.4 GHz, nonthermal radio component,
thermal radio component, and
CO~(2-1). The
arrows at $\rm r=0.1$~kpc show the 3$\sigma$ upper-limits of CO~(6-5)
and CO~(2-1) in the central hole.
}
\label{fig:comp_co65_10_radio}
\end{figure}

In Figure~\ref{fig:comp_co65_10_radio} the radial profiles
of the emissions are compared. 
The peak of the radial distribution of the
thermal radio is
shifted (by $\sim 70$~pc) toward the smaller radius compared to that of
the CO~(6-5) radial distribution. 
The peak of the distribution of the total radio emission also has a 
small offset compared to that of the CO~(6-5), while 
the radial profile of nonthermal radio emission 
is similar to that of the CO~(6-5)
in the ring region.
The radio profile of CO~(2-1) \citep{Konig2013} is also shown in the same plot.
If we compare the CO~(2-1) profile
with the profile of CO~(6-5), which has been smoothed to the same 
resolution of CO~(2-1), we see that
the former is significantly more extended than 
the latter. Beyond $\rm r \sim 400\; pc$,
the cold and diffuse molecular gas probed by CO~(2-1) emission 
is devoid of any significant star formation (as revealed
by the profiles of the radio emissions).

\begin{figure}[!htb]
\epsscale{1.2}
\plotone{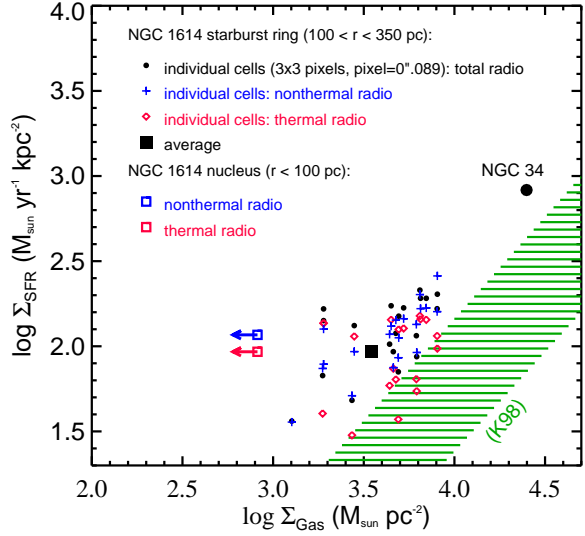}
\caption{
Plot of $\rm log \Sigma_{SFR}$ versus $\rm log \Sigma_{Gas}$.
For individual cells in the NGC~1614 ring: 
$\rm log \Sigma_{SFR,th}$ (red diamonds) and $\rm log \Sigma_{SFR,nth}$ 
(blue crosses), and $\rm log \Sigma_{SFR,total}$ (black dots)
are estimated using the thermal, nonthermal, and total radio maps, 
respectively; and $\rm log \Sigma_{Gas}$ is estimated from the CO~(6-5) map 
that is smoothed and regridded to match the radio maps. 
For the NGC~1614 nucleus (open squares): the 3-$\sigma$ 
upper-limit for $\rm \Sigma_{Gas}$ was derived using the CO~(6-5) map
assuming the same relation for the ring region (Eq~\ref{eq:gas}).
The average for the NGC~1614 ring  (black solid square): 
data taken from Table~\ref{tbl:comparison}.
Nuclear starburst in NGC~34 (black solid circle): data taken from
Table~\ref{tbl:comparison}. The shaded area (in green color) 
represents the data for
local starbursts in the sample of \citet{Kennicutt1998b}.
}
\label{fig:ksplot}
\end{figure}

In Figure~\ref{fig:ksplot} we plot the SFR surface density 
($\rm \Sigma_{SFR}$) vs. the gas surface density ($\rm \Sigma_{Gas}$) 
(i.e. the Kennicutt-Schmidt law) for
the nuclear starburst and individual cells ($\rm 3\time 3$ pixels,
pixel=0\farcs{089}) in the ring, using the
thermal and nonthermal maps to derive $\rm \Sigma_{SFR}$
and the CO~(6-5) map (smoothed and regridded to match the radio maps)
to obtain $\rm \Sigma_{Gas}$.
The SFR can be estimated from the nonthermal and thermal radio luminosities 
using two formulae given in \citet{Murphy2012}, respectively:
\begin{equation}
\rm 
\left( {SFR^{nth}_\nu\over M_\sun\; yr^{-1}} \right) =
6.64\times 10^{-29}\left( {\nu\over GHz}\right)^{\alpha^{nth}}\left(
{L^{nth}_\nu\over erg\; s^{-1}Hz^{-1}}\right). \label{eq:nth}
\end{equation}
and 
\begin{eqnarray}
\rm 
\left( {SFR^{th}_\nu\over M_\sun\; yr^{-1}} \right) & = &
\rm 4.6\times 10^{-28}\left({T_e\over 10^4 \; K}\right)^{-0.45} 
\times \nonumber  \\
& &\rm \left({\nu\over GHz}\right)^{0.1} \left({L^{th}_\nu\over erg\; s^{-1}\; Hz^{-1}}\right).
\end{eqnarray}
where $\rm T_e= 10^4 \rm K$, $\nu = 8.4\; GHz$, and $\rm  \alpha^{nth}=1.2$ 
\citep{Herrero-Illana2014}.
For  individual cells in the ring,
we also plotted in Figure~\ref{fig:ksplot}
the $\rm \Sigma_{SFR}$ vs. $\rm \Sigma_{Gas}$ relation with
the $\rm \Sigma_{SFR}$ estimated from the total radio emission, assuming
a constant nonthermal fraction ($\rm f_{nth}=0.5$) and the SFR vs. $\rm L^{nth}$
relation in Eq~\ref{eq:nth}.

The gas surface density 
was estimated using the CO~(6-5) surface brightness
as following:  According to
\citet{Konig2013}, the total $\rm H_2$ mass 
in the ring is $\rm  M_{H_2} = 10^{8.97}\; M_\sun$ 
(for D=67.8~Mpc), estimated using the CO~(1-0) map of \citet{Olsson2010}
and assuming a conversion factor of $\rm 3\times 10^{20}\; cm^{-2}\; 
(K\; km\; s^{-1})^{-1}$. Dividing this
by the integrated CO~(2-1) flux of the ring, $\rm S_{CO(2-1)} = 65.4\pm 6.9\;
Jy\; km\; s^{-1}$ \citep{Konig2013} and  
assuming a brightness temperature ratio of 0.72 
between CO~(6-5) and CO~(2-1) (Figure~\ref{fig:compCO21}), 
we have: 
\begin{equation}
\rm 
\left( {\Sigma_{Gas}\over M_\sun\; pc^{-2}} \right) = 20.3\times 
\left( {f_{CO(6-5)}\over Jy\; arcsec^{-2} \; km\; s^{-1}} \right). \label{eq:gas}
\end{equation}

In the ring region, only cells that are detected 
in both radio and CO~(6-5) maps above a 3-$\sigma$ threshold are plotted
(therefore the random errors are $\rm <0.12$ dex for these data points). 
In the nuclear region ($\rm r < 100$~pc), the 3-$\sigma$ 
upper-limit for $\rm \Sigma_{Gas}$ was derived using the CO~(6-5) map
assuming the same relation for the ring region (Eq~\ref{eq:gas}).
For individual cells in the ring, the $\rm \Sigma_{SFR}$ 
vs.  $\rm \Sigma_{Gas}$ relation
is systematically above that for local starbursts
\citep{Kennicutt1998b}, indicating a higher star formation efficiency
(SFE).  This is because, by relating the SFR to the warm dense gas
probed by the high resolution ALMA observations of CO~(6-5), much of
the cold diffuse gas probed by low J CO (more extended than the warm
dense gas) is excluded from the $\rm \Sigma_{Gas}$ in our results. 
It is worth noting that we used a standard CO conversion factor ($\rm
X_{CO} = 3\times 10^{20}\; cm^{-2}(K\; km\; s^{-1})^{-1}$) for
NGC~1614 data. In the literature, arguments for high SFE in (U)LIRGs
are very often based on results obtained using a CO conversion factor
$\sim 5$ times lower than the standard value \citep[e.g.,][]{Daddi2010,
  Genzel2010}.

It appears that, on  the linear scale of
100 pc, the tight correlation previously found 
between $\rm \Sigma_{SFR}$ vs. $\Sigma_{Gas}$ \citep{Kennicutt1998b,
Genzel2010, Leroy2013, Yao2003, Iono2004, Wilson2008}
breaks down in the central kpc of NGC~1614. In particular,
the non-detections of the nucleus in
both CO~(6-5) and CO~(2-1) maps
set a lower-limit of the
$\rm \Sigma_{SFR}$-to-$\Sigma_{Gas}$ ratio about an order of magnitude above
the nominal value, corresponding to a very short gas exhaustion
time scale of $\rm M_{gas}/SFR < 10$ Myr.
The low extinction \citep{Alonso-Herrero2001, Kotilainen2001,
Diaz-Santos2008} and low PAH emission \citep{Vaisanen2012} also 
indicate an ISM depression in the nuclear region.
The star formation time
scale associated with the thermal radio is $\sim 10$ Myr
and that with the nonthermal radio is $\sim 100$ Myr.
\citet{Alonso-Herrero2001} argued that, 
based on detections of deep CO stellar absorption, 
NGC~1614 harbors a nuclear starburst older than 10 Myr, which could
have blown away the ambient ISM \citep{Vaisanen2012}.
If the time scale for the feedback effects, including
both the gas consumption by star formation and mass loss by
superwinds, is significantly shorter than 10 Myr
(the dynamic time scale of the nuclear region is only
$\rm \tau \sim 1$ Myr), then the deviation of the nucleus
from the $\rm \Sigma_{SFR}$ vs. $\Sigma_{Gas}$ relation 
may indeed be due to the feedback of the old nuclear starburst.
This is consistent with the results of \citet{Garcia-Burillo2012}
who found that NGC~1614 has the highest value of the SFE 
(estimated from the FIR/HCN ratio) among a sample of normal star-forming 
galaxies and mergers, and argued that this could be due to  
the exhaustion of the dense molecular gas by starburst activity.

In the starburst ring, the correlation between 
$\rm \Sigma_{SFR,th}$ and $\Sigma_{Gas}$ is rather weak
(Spearman's rank correlation coefficient $\rm \rho =0.37$ with
the significance of its deviation from zero $\rm p = 0.20$).
Also, the correlation between 
$\rm \Sigma_{SFR,total}$ (estimated using total radio emission)
and $\Sigma_{Gas}$  has a large scatter, and is only marginally significant 
($\rm \rho =0.64$ and $\rm p = 0.0023$). This is consistant with  
the systematic offset between the radial profiles of
the total radio and CO~(6-5) in Figure~\ref{fig:comp_co65_10_radio}. 
While the weak correlation between 
$\rm \Sigma_{SFR,th}$ and $\Sigma_{Gas}$ could be mainly due to the 
uncertainties associated with 
the obscuration correction of the $\rm Pa\; {\alpha}$ (the thermal radio is
estimated using the obscuration-corrected $\rm Pa\; {\alpha}$), this
cannot explain the lack of correlation between $\rm \Sigma_{SFR,total}$ 
and $\Sigma_{Gas}$.
We have already seen a breakdown of the $\rm \Sigma_{SFR}$-to-$\Sigma_{Gas}$
correlation in the nucleus, and interpreted it as a consequence 
of starburst feedback. The same scenario can be applied to the individual
cells in the ring. Given the high resolutions of the radio and CO~(6-5)
maps, these cells correspond to star formation regions of linear scales
of $\sim 100~pc$. On such fine scales, the $\rm \Sigma_{SFR}$-to-$\Sigma_{Gas}$
relation could be sensitive to the local star-formation history.
Indeed, \citet{Alonso-Herrero2001} suggested that in NGC~1614
the starburst propagates like a ``wild fire'' from the nucleus
outward. \citet{Vaisanen2012} proposed that even the ring is stratified
in terms of the star formation age. In a CO~(1-0) survey of
M33, \citet{Onodera2010} found
a breakdown of the Kennicutt-Schmidt law 
on the linear scale of $\rm \sim 80~pc$,
and attributed it to the various evolutionary stages of 
giant molecular clouds (GMCs) and to the drift of young clusters from their
parent GMCs. These interpretations are applicable to our results, although
our ALMA observations probe an even tighter correlation between
CO~(6-5) and SFR in a LIRG associated with a starburst merger.

A stronger correlation is found between $\rm
\Sigma_{SFR,nth}$ and $\rm \Sigma_{Gas}$ in the starburst ring
($\rm \rho =0.81$ and $\rm p = 1.6\times 10^{-5}$).
This is puzzling because, given
the longer star formation time scale associated with the nonthermal
radio, this relation should be more sensitive to the star formation
history than the $\rm \Sigma_{SFR,th}$ and $\rm \Sigma_{Gas}$
relation.  It is likely that the $\rm \Sigma_{SFR,nth}$ and $\rm
\Sigma_{Gas}$ correlation is driven by other factors than the
Kennicutt-Schmidt law.  One possibility is that it is due to the
correlation between the magnetic field strength and the gas density
\citep{Fiebig1989, Helou1993, Niklas1997}. Observationally, this
correlation extends from the smallest \citep{Fiebig1989} to the
largest cosmic scales \citep{Vallee1990, Vallee1995}, and has the form of $\rm B
\propto n^{k}$ for $\rm n > 10^2\; cm^{-3}$, where B is the magnetic
field strength, n the gas density, and $\rm k = 0.5\pm 0.1$
\citep{Fiebig1989}. Since the
emissivity of the nonthermal (synchrotron) radiation is proportional
to $\rm B^2$, the B vs. n correlation leads naturally to a localized
($\sim$ linear) correlation between nonthermal surface brightness and
gas surface density. Another possibility is that the nonthermal radio
and the CO~(6-5) correlate with each other because they are both
powered by cosmic rays (CRs).  Indeed, CSO observations of $\rm
^{12}$CO(6-5) and $\rm ^{13}$CO(6-5) by \citet{Hailey-Dunsheath2008}
of the nuclear starburst in the central 180 pc of NGC~253 suggested
that warm molecular gas is most likely to be heated by an elevated
density of CRs or by turbulence.  In order to test whether CRs
dominate the heating of warm molecular gas in NGC~1614, we carried out
model-fitting using theoretical models \citep{Meijerink2005,
Kazandjian2012, Kazandjian2014} of the
Cosmic Ray Dominated Regions (CDRs) and Photon Dominated Regions
(PDRs) to fit the emission lines of $^{12}$CO, $^{13}$CO, HCN, HNC,
and HCO$^+$. These are mostly single dish data for the entire system of
NGC~1614 \citep{Sanders1991, Albrecht2007, Costagliola2011, Lu2014b},
plus some high resolution interferometry data for the central region
taken from the literature \citep{Wilson2008, Olsson2010, Konig2013,
  Imanishi2013} and from this work.  The results show that PDR models
with strong mechanical heating (by turbulence) provide the best fit
while CDR models fit the data rather poorly. Details of these results
will be presented elsewhere (Meijerink et al., in preparation).  This
is consistent with \citet{Rosenberg2014a} who modeled the Herschel
observations of $\rm ^{12}$CO up to upper J=13 and $\rm ^{13}$CO up to
upper J=6, together with data of other submm lines taken from the
literature, of NGC~253. They found that mechanical heating by
turbulence is necessary to reproduce the observed molecular emission
and CR heating is a negligible heating source. \citet{Rosenberg2014b}
reached a similar conclusion for Arp~299A, a nuclear starburst in 
Arp~299 (a merger-induced LIRG).  In principle, the turbulence can be
related to the CRs through shocks generated by supernova remnants
(SNRs) which can both power the turbulence \citep{Draine1980} and
accelerate CRs \citep{Drury1994}.  However, given the very different
mechanisms for energizing low velocity turbulence and for CR
acceleration by SNR shocks, it is unlikely that this can explain the
localized correlation between $\rm \Sigma_{SFR,nth}$ and $\rm \Sigma_{Gas}$ in
the starburst ring down to the linear scale of 100 pc.

\subsection{NGC~1614 and NGC~34: A Tale of Two LIRGs}

\begin{deluxetable}{lccc}
\tabletypesize{\normalsize}
\setlength{\tabcolsep}{0.03in} 
\tablecaption{Comparison between NGC~1614 and NGC~34 \label{tbl:comparison}
}
\tablehead{
      & NGC~1614 & & NGC~34
}
\startdata
R.A. (J2000)$\rm ^a$ & $\rm 04{\hr}34{\mn}00{\fs}03$ & &  $\rm 00{\hr}11{\mn}06{\fs}54$ \\
Dec. (J2000)$\rm ^a$ & $\rm -08{\degr}34{\arcmin}45{\farcs}1$&&$\rm -12{\degr}06{\arcmin}27{\farcs}5$ \\ 
Distance (Mpc) & 67.8 & & 84.1 \\
$\rm L_{IR}$ ($\rm L_\sun$)$\rm ^b$ & $10^{11.65}$ & &$10^{11.49}$ \\
$\rm M_{K}$ (mag)$\rm ^c$&  $-24.59$ & & $-24.46$  \\
$\rm M_{HI}$ ($\rm M_\sun$)$\rm ^d$ & $10^{9.45}$ && $10^{9.72}$  \\
$\rm SFR_{tot}\; (M_\sun\; yr^{-1}$)$\rm ^e$  & 51.3 & & 34.7 \\
$\rm M_{H_2, tot}$ ($\rm M_\sun$)$\rm ^f$&  $10^{10.12}$ && $10^{10.15}$  \\
$\rm M_{dust,tot}$ ($\rm M_\sun$)$\rm ^g$&  $10^{7.60}$ && $10^{7.48}$  \\
AGN &  No & & Yes  \\
Merger mass ratio & 4:1 -- 5:1 & & 3:2 -- 3:1  \\
$\rm S_{8.4GHz, tot}$ (mJy)$\rm ^h$ & 41.1 & & \\
$\rm S_{CO~(6-5), tot}$ ($\rm Jy\; km\; s^{-1}$)$\rm ^i$ & $1423\pm 126$ & & $937\pm 63$ \\
$\rm S_{435\mu m, tot}$ (mJy)$\rm ^j$ & $831\pm 58$ & & $517\pm 36$  \\
& & & \\
{\bf Central Starburst:} & & & \\
Morphology & circum-nuclear ring  & & nuclear disk \\
radius (pc) & $\rm r_{in}=100,\; r_{out}=350$ & & $\rm 100$ \\  
$\rm S_{8.4GHz, cent}$ (mJy)$\rm ^k$ & 26.5 & & 15.2 \\
$\rm SFR_{cent}\; (M_\sun\; yr^{-1}$)$\rm ^l$  & 32.8 & & 26.0 \\
$\rm \Sigma_{SFR}$ ($\rm M_\sun\; yr^{-1}\; kpc^{-2}$)$\rm ^m$  & 92.8 & & 
827.6 \\
$\rm M_{H_2, cent}$  ($\rm M_\sun$)$\rm ^n$ & $10^{8.97}$ & &  $10^{8.76}$ \\
$\rm \Sigma_{Gas}$ ($\rm M_\sun\; pc^{-2}$)$\rm ^o$  & $10^{3.54}$ & & $10^{4.40}$ \\
$\rm S_{CO~(6-5), cent}$ ($\rm Jy\; km\; s^{-1}$)$\rm ^p$ & $898\pm 153$ & & $1004\pm 151$ \\
$\rm S_{435\mu m, cent}$ (mJy)$\rm ^q$ & $269\pm 46$ & & $275\pm 41$  \\
$\rm M_{dust, cent}$  ($\rm M_\sun$)$\rm ^r$ & $10^{7.11}$ & &  $10^{6.97}$ \\
\enddata
\tablecomments{{\small{
\\
{$\rm ^a$} Coordinates of the nucleus in the 8.4 GHz radio continuum.\\
{$^b$} IR luminosity between 8 -- 1000 $\mu m$ 
      \citep{Armus2009}.\\
{$^c$} Absolute K band magnitude \citep{Rothberg2004}.\\
{$^d$} Total mass of neutral atomic hydrogen gas, taken from
compilation by \citep{Kandalyan2003}.\\
{$^e$} Total star formation rate \citep{U2012}.\\
{$^f$} Total mass of molecular hydrogen gas (assuming
$\rm X_{CO} = 3\times 10^{20}\; cm^{-2}(K\; km\; s^{-1})^{-1}$); 
NGC~1614: \citet{Sanders1991}; NGC~34: \citet{Krugel1990}. \\
{$^g$} Total dust mass; NGC~1614: this work; NGC~34: \citet{Esquej2012}.\\
{$^h$} Total flux of the 8.4 GHz radio continuum \citep{Schmitt2006}.\\
{$^i$} Total flux of the CO~(6-5) emission \citep{Lu2014b}.\\
{$^j$} Total flux of the 435$\mu m$ continuum emission;
NGC~1614: this work; NGC~34: \citet{Xu2014}.\\
{$^k$} Flux of the 8.4 GHz radio continuum in the central region;
NGC~1614: \citet{Herrero-Illana2014}; NGC~34: \citet{Condon1991}.\\
{$^l$} The SFR of central starburst:
$\rm SFR_{cent} = SFR_{tot}\times f_{cent}$,
where 
$\rm f_{cent} = S_{8.4GHz,cent}/S_{8.4GHz,tot} = 0.64$ for NGC~1614, and 
$\rm f_{cent} = 0.75 $ for NGC~34.\\
{$^m$} Mean SFR column density of the central starburst.\\
{$^n$} Mass of molecular hydrogen gas in the central region;
NGC~1614 ($\rm X_{CO} = 3\times 10^{20}\; cm^{-2}(K\; km\; s^{-1})^{-1}$): 
\citep{Konig2013};
NGC~34 ($\rm X_{CO} = 0.5\times 10^{20}\; cm^{-2}(K\; km\; s^{-1})^{-1}$): 
\citet{Fernandez2014}.\\
{$^o$} Mean gas column density of the central starburst 
($\rm M_{Gas} = 1.36\times M_{H_2}$).\\
{$^p$} Flux of the CO~(6-5) emission in the central starburst region;
NGC~1614: this work; NGC~34: \citet{Xu2014}.\\
{$^q$} Flux of the 435$\mu m$ continuum emission in the central starburst region;
NGC~1614: this work; NGC~34: \citet{Xu2014}.\\
{$^r$} Dust mass in the central starburst region;
NGC~1614: this work; NGC~34: \citet{Xu2014}.\\
}}}
\end{deluxetable}


In this section we compare NGC~1614 with NGC~34,
another local LIRG observed by
our team using ALMA band-9 receivers \citep{Xu2014}.
Both galaxies are late-stage
mergers \citep{Neff1990, Schweizer2007}.
As shown in Table~\ref{tbl:comparison},
they have similar absolute K-band magnitude $\rm M_K$ (indicating
similar stellar mass),
similar total gas mass as obtained by HI and CO observations,
and similar total SFR as derived from the IR+UV luminosities
\citep{U2012}. On the other hand, as revealed by the ALMA observations
and high angular resolution observations in other bands, 
the two galaxies are very different in the central kpc.

First of all, our ALMA data ruled out a Compton-thick
AGN in NGC~1614. By comparison, there is a weak AGN in NGC~34
according to the X-ray data \citep{Brightman2011a, Esquej2012},
and the ALMA results \citep{Xu2014} are consistent with the AGN being
Compton thick. Nevertheless, for both galaxies, the central kpc is
dominated by starburst activity and AGN contributions to
both dust and gas heatings are insignificant 
\citep{Xu2014, Stierwalt2013}. In NGC~34, the starburst is
concentrated in a compact nuclear disk of $\rm r \sim 100\; pc$, 
with very high $\rm \Sigma_{SFR,th}$ and 
$\Sigma_{Gas}$. In NGC~1614, a starburst ring
between $\rm r_{in}=100\; pc$ and $\rm r_{out}=350\; pc$ dominates the central
region, with moderate mean $\rm \Sigma_{SFR,th}$ and mean
$\Sigma_{Gas}$ compared to other local starbursts (Figure~\ref{fig:ksplot}).
It is worth pointing out that different CO conversion factors have
been adopted for the two cases: For the nuclear starburst in NGC~34, 
the ALMA observations showed that the molecular gas is concentrated in
a well organized disk controlled 
mostly by the gravity of stars \citep{Xu2014}.
Therefore, we choose to use the conversion factor for (U)LIRGs:
$\rm X_{CO} = 0.5\times 10^{20}\; cm^{-2}(K\; km\; s^{-1})^{-1}$
\citep{Scoville1997, Downes1998}. 
On the other hand, for
the starburst ring in
NGC~1614,
the ALMA observations presented here and the SMA observations of 
\citet{Konig2013}
reveal that much of the CO emission is clumped in individual knots 
associated with giant molecular associations (GMAs), which might be 
self-gravitating. 
In this case, a standard Galactic CO conversion factor is more 
appropriate \citep{Papadopoulos2012a}:
$\rm X_{CO} = 3.0\times 10^{20}\; cm^{-2}(K\; km\; s^{-1})^{-1}$.
Nevertheless, these conversion factors are very uncertain and are 
the major error sources for the molecular gas mass estimates.

\begin{figure}[!htb]
\plotone{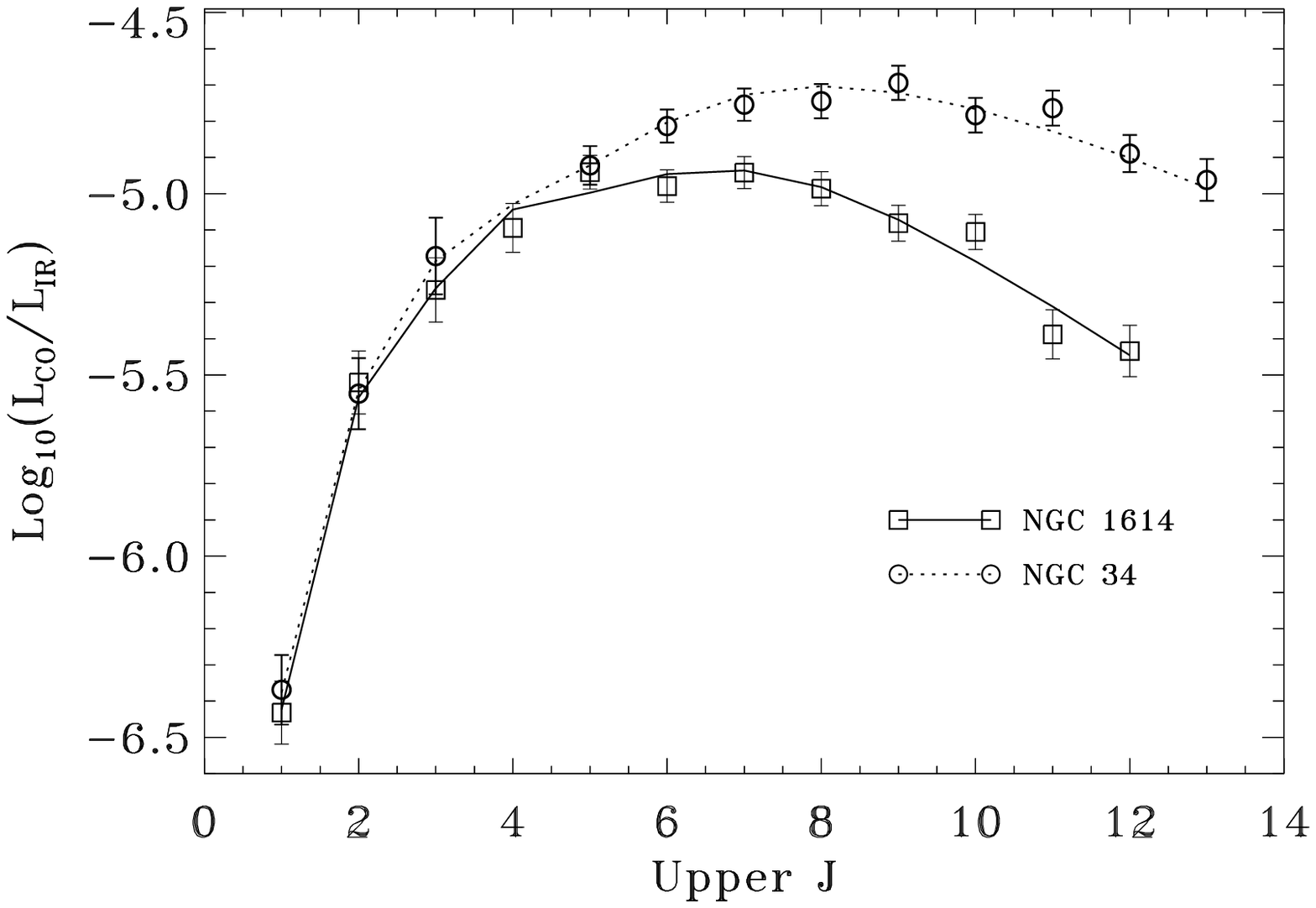}
\caption{Plot of the $\rm L_{IR}$ normalized 
spectral line energy distributions (SLEDs) of
NGC~1614 and NGC~34.
The data points (all obtained by single-dish observations) 
are taken from the literature with the following references:
for NGC~1614: \citet{Sanders1991} for CO~(1-0); 
\citet{Albrecht2007} for CO~(2-1); \citet{Wilson2008} 
for CO~(3-2); \citet{Lu2014b} for CO~(4-3) and other higher-J lines;
for NGC~34: \citep{Albrecht2007} and \citet{Maiolino1997} for CO~(1-0);
\citet{Papadopoulos1998} for CO~(2-1); 
Zhang et al. (2014, in preparation) for CO~(3-2); \citet{Lu2014b} for CO~(4-3) 
and other higher-J lines. The solid (dotted) line is model fitting of the
CO SLED of NGC~1614 (NGC~34).}
\label{fig:co-sled}
\end{figure}

In Figure~\ref{fig:co-sled} we compare the spectral line energy
distributions (SLEDs) of the total CO emission (measured by
single-dish observations) of these two galaxies, taken from
observations of Herschel SPIRE FTS observations \citep{Lu2014b}.  The
CO SLED of NGC~1614 peaks around upper J = 5-7, while that of NGC~34 reaches a
plateau after a rapid increase, and the peak is around upper J=9.  In order
to further investigate the physical conditions of these two galaxies, we
modeled the observed CO SLEDs using simple two-component
RADEX large velocity gradient (LVG) radiative
transfer models \citep{vanderTak2007}, and adopting a similar
procedure to that in \citet{Kamenetzky2012}. 
Admittedly, results from such model fittings
suffer significant degeneracy between parameters \citep{Rosenberg2014a}.
Nevertheless, they are useful for translating the information 
in the CO SLED into quantitative estimates of physical parameters of the gas,
albeit with large uncertainties.  We find that both SLEDs
can be well fitted by the combination of a cool and a warm
component. Both galaxies have similar gas densities of ~($10^{2.5}$,
$10^{2.6}$) cm$^{-3}$ and ($10^4$, $10^4$) cm$^{-3}$, for the cool and
warm components in (NGC~34, NGC~1614), respectively. However, the
kinetic temperature of the warm component in NGC~34 (890 K) is 2 times
higher than that of NGC~1614 (445 K), consistent with the fact that
the nuclear starburst in NGC~34 is 5 times more compact than the
circum-nuclear starburst ring in NGC~1614. It is worthwhile noting
that the AGN contribution to the warm gas in NGC~34 is insignificant
\citep{Xu2014}. Given the overall similarities between the two host galaxies 
(Table~\ref{tbl:comparison}), it is likely that
the staunch difference between the two central starbursts is 
caused by the difference 
in the merging processes that the two LIRGs have experienced.

The morphology of NGC~1614---one prominent tail and one relatively
small secondary tail---suggests an unequal mass encounter (mass ratio
$\gtrsim 4:1$) and/or a scenario in which one of the galaxies
experienced a retrograde passage. Several authors have argued for a
high mass ratio encounter; \citet{Rothberg2006} note the isophotal
shape of NGC 1614 and its correspondence with simulations of high mass
ratio mergers, while \citet{Vaisanen2012} identify a possible remnant
body of the lower-mass companion. Both \citet{Rothberg2006} and
\citet{Vaisanen2012} come to the same conclusion---that NGC 1614 is a
4:1 mass ratio merger---but the former assumes the nuclei have already
merged and the latter relies on the identification of an interacting
galaxy.

NGC~34 has no clear evidence for dual nuclei, suggesting the two
galaxies have already coalesced. Owing to the asymmetry of integrated
brightness of the two tidal tails \citet{Schweizer2007} argue this
system is the result of a merger of two disk galaxies with a mass
ratio between 3:2 and 3:1. The disky isophotal shape of the remnant
(which shows no evidence for a disk in the K-band morphology) is
consistent with a formation scenario of a major but unequal mass
merger \citep{Rothberg2006,Naab2006}. Preliminary dynamical modeling
of this system is consistent with the aforementioned mass ratio and
both disks experiencing prograde interactions (G. C. Privon et
al.\ \emph{in prep}). This dynamical model is consistent with the
system being observed $\sim250-300$ Myr since the first passage of the two
galaxies, somewhat lower than the suggested $400$ Myr age of the
stellar disk \citep{Schweizer2007}.

Hence, NGC~34 has experienced a major merger of two galaxies of
similar mass, which was catastrophic and destroyed both
progenitor disks \citep{Schweizer2007}. Simulations by \citet{Cox2008}
exploring the effect of mass ratio on merger-induced starbursts found
a decreasing burst strength with increasing primary/secondary mass
ratio; given the previously mentioned estimates of the mass ratios for
NGC 34 and NGC 1614, the star formation surface densities are
consistent with this interpretation. It might be that the higher mass
ratio merger experienced by NGC~1614 caused less efficient torquing of
the gas, leading to much of the central gas settling into the nuclear ring (with
the help of either the inner Lindblad resonance associated with a bar
\citep{Olsson2010} or the non-axisymmetric potential caused by a minor merger
\citep{Combes1988, Knapen2004, Mazzuca2006}) rather
than collecting in the center, as in NGC~34. This may also hint
at the answer to the question why NGC~1614 has not yet developed an
AGN \citep{Vaisanen2012} while NGC~34 has one.  According to
\citet{Hopkins2012a}, the built-up of a centrally peaked dense gas
disk is a necessary condition for triggering of the AGN activity in
late stage mergers.

An alternate explanation for NGC~1614's comparatively lower $\rm
\Sigma_{SFR}$, if the scenario proposed by \citet{Vaisanen2012} is
accurate, is that the merger has simply not yet run to completion and
so has not yet caused the final funneling of gas towards the nucleus
at the time of the merger \citep[e.g.,][]{Mihos1994b,
  Hopkins2012a}. \citet{Olsson2010} and \citet{Konig2013} both 
show that indeed most of the molecular gas in NGC~1614
sits in the dust lane (outside the ring) and even further out. 
It could be that the relatively minor perturbation of
the first pass (which led to the northeast tail) created 
the outward propagating starburst (i.e. the ``wild fire''), as revealed
by the nuclear ring and the weak and old nuclear starburst,
while a future merger will trigger a much stronger
nuclear starburst as seen in NGC~34.

With current knowledge of the encounters in NGC~34 and NGC~1614, we
cannot firmly assign the cause for the different starburst
characteristics in the two systems. It is likely to be
due to the effect of different mass ratios, 
but we cannot rule out other possible causes such as
different current phases of the encounters and
different encounter geometries. 
While NGC~34 represents a large population of LIRGs with
starburst nuclei (e.g. Arp~220),  NGC~1614 represents those with
circum-nuclear starburst rings, which are also common in LIRGs.
Among the GOALS sample, at least five other LIRGs (NGC~1068, NGC~5135, 
NGC~7469, NGC~7552, and NGC~7771) have such rings.
Future dynamical models (G. C. Privon et al.\ \emph{in
  prep}) matched to the
kinematics and morphology of NGC~34 and NGC~1614 may
provide a more concrete answer to the question of how the two galaxies, 
and the two LIRG populations they represent, developed such different
central starbursts over the merging process.

\section{Summary}\label{sect:summary}
We carried out ALMA observations of the CO~(6-5) line emission and of
the 435~$\mu m$ dust continuum emission in the central kpc of
NGC~1614, a local LIRG at distance of 67.8
Mpc ($\rm 1\arcsec = 329\; pc$).  The CO emission and the continuum
are both well resolved by the ALMA beam ($\rm 0\farcs26\times
0\farcs20$) into a circum-nuclear ring. The integrated flux of
CO~(6-5) is $\rm f_{CO~(6-5)} = 898\; (\pm 153) \; Jy\; km\; s^{-1}$,
and the flux of the continuum is $\rm f_{CO~(6-5)} = 269\; (\pm 46)
mJy$. These are $\rm 63(\pm 12) \%$ and $\rm 32(\pm 6) \%$ of the
total CO~(6-5) flux and 435~$\mu m$ continuum flux of NGC~1614
measured by Herschel, respectively. The molecular ring, located
between $\rm 100\; pc < r < 350\; pc$, looks clumpy and
includes several unresolved (or marginally resolved) knots with median
velocity dispersion of $\rm \delta v \sim 40\; km\; s^{-1}$.  These
knots are associated with star formation regions with $\rm
\Sigma_{SFR}\sim 100\; M_\sun\; yr^{-1}\; kpc^{-2}$ and $\rm
\Sigma_{Gas}\sim 10^4\; M_\sun\; pc^{-2}$. The non-detections of the
nucleus in both the CO~(6-5) and the 435 $\mu m$ continuum rule out,
with relatively high confidence, a Compton-thick AGN in NGC~1614.
Comparisons with the radio continuum show that the local correlation,
on the linear scale of $\sim 100$~pc, between $\rm \Sigma_{Gas}$ and
$\rm \Sigma_{SFR}$ (i.e. the Kennicutt-Schmidt law) is severely
disturbed. In particular, the nucleus has a lower-limit of the $\rm
\Sigma_{SFR}$-to-$\Sigma_{Gas}$ ratio about an order of magnitude
above the nominal value in the standard Kennicutt-Schmidt law.  This
break-down of the star formation law could be caused by an outward
propagation of the central starburst (i.e. the ``wild fire''
scenaio proposed by \citealt{Alonso-Herrero2001}). 
Our results also show that the CO~(6-5) correlates stronger
with the nonthermal radio component than both
the total radio emission and the thermal radio component, 
possibly due to an in situ correlation between the magnetic
field strength and the gas density.


\vskip1truecm
\noindent{\it Acknowledgments}:
Adam Leroy and Tony Remijan from NAASC are thanked for their help with
data reduction. An anonymous referee is thanked for constructive comments. 
Y.G. is partially supported by NSFC-11173059,
NSFC-11390373, and CAS-XDB09000000. Y.Z. thanks the NSF of Jiangsu
Province for partial support under grant BK2011888. V.C. would like to
acknowledge partial support from the EU FP7 Grant
PIRSES-GA-2012-316788. This paper makes use of the following ALMA
data: ADS/JAO.ALMA-2011.0.00182.S. ALMA is a partnership of ESO
(representing its member states), NSF (USA) and NINS (Japan), together
with NRC (Canada) and NSC and ASIAA (Taiwan), in cooperation with the
Republic of Chile. The Joint ALMA Observatory is operated by ESO,
AUI/NRAO and NAOJ.  This research has made extensive use of the
NASA/IPAC Extragalactic Database (NED) which is operated by the Jet
Propulsion Laboratory, California Institute of Technology, under
contract with the National Aeronautics and Space Administration.

\bibliographystyle{apj}
\bibliography{/Volumes/Seagate/data1/bibliography/papers_biblio}

\appendix

\section{Thermal/Nonthermal Decomposition of the Radio Continuum}

The thermal radio emission is estimated
from an extinction-corrected Pa-$\alpha$ map of the central
region of NGC1614 derived from HST NICMOS imaging data 
\citep{Alonso-Herrero2001} 
taken in four NIR continuum and emission line bands
(F160W and F222M, and F187N and F190N respectively). The maps are 
background-subtracted \citep{Diaz-Santos2008}. The
dust-attenuated Pa-$\alpha$ emission map was obtained from the F190N
narrow-band line+continuum image after the subtraction of the adjacent
continuum emission obtained with the F187N filter. In order to correct
the Pa-$\alpha$ map for the significant extinction 
(e.g., \citealt{Neff1990, Alonso-Herrero2001, Kotilainen2001, Rosenberg2012}),
we used the equivalent width (EW) map of the emission line
plus two broad-band NIR continuum images obtained at 1.6 and
2.2um.

First, we used Starburst99 (v7.0.0; \citealt{Leitherer1999}) to
generate a stellar population synthesis model for an instantaneous
burst of star formation with Geneva evolutionary stellar tracks,
Kroupa initial mass function, and solar metallicity. The model outputs
were obtained with a 0.1 Myr step for starburst ages ranging from 0.01
to 50 Myr. In addition to the spectral energy distribution (SED) of
the continuum, the model also provides the nebular emission from
hydrogen recombination lines. Assuming that the extinction to the gas
is similar to that towards the dust, the observed Pa-$\alpha$ EW map was
compared to the model predictions to estimate the age of the young
stellar population. We note that the ages are upper limits to the real
ages since the presence of an older, underlying stellar population
would increase the NIR continuum, thus aging the regions.

\begin{figure*}[!htb]
\plotone{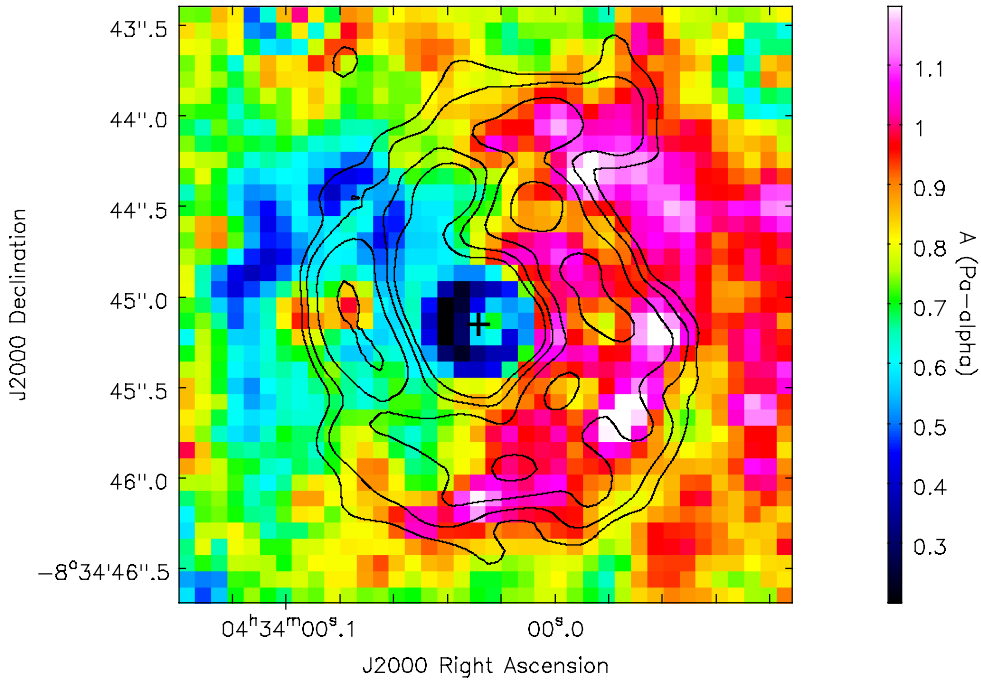}
\vskip-10cm
\caption{Extinction map in units of $\rm A_{Pa-\alpha}\; (mag)$
constructed based on the comparison between the observed NIR color F160W/F222M and that derived from Starburst99 models for the stellar population ages estimated from the Pa-$\alpha$ EW map. Contours of the CO~(6-5) line emission
are overlaid on the map.
}
\label{fig:NIRextinction}
\end{figure*}

Once the ages of
the stellar populations are estimated, we can compare the NIR continuum
slope derived from the two continuum broad-band
filter images with the model SEDs to derive the obscuration. To this
end, the Starburst99 spectra were convolved for each age step with the
corresponding continuum filters. The synthetic NIR colors 
(F160W/F222M) were interpolated to the ages derived
from the Pa-$\alpha$ EW map and compared with the observed values,
and the extinction was estimated from this comparison.
The dust attenuation law used to derive the
extinction was that of \citet{Calzetti2000} for a foreground dust screen
configuration. The resulting extinction map in units of 
$\rm A_{Pa-\alpha}\; (mag)$ is shown Figure~\ref{fig:NIRextinction}.
The typical error of $\rm A_{Pa-\alpha}\; (mag)$ is 0.3 mag, estimated from
the uncertainties of the age and the color map. 
The strong east-west asymmetry in the $\rm A_{Pa-\alpha}$ map is in agreement
with that seen in the low-J CO maps \citep{Olsson2010, Konig2013}, and is 
likely associated with a broad dust-lane extended from the north to the west
of the ring \citep{Konig2013}. 
Finally, the observed Pa-$\alpha$ image was corrected
using the obscuration map obtained from this method. The
extinction-corrected Pa-$\alpha$ emission was then scaled to obtain the
thermal component of the radio emission at 8.4 GHz (see equation 4 in
\citealt{Herrero-Illana2014}). The nonthermal component was then
obtained by subtracting the thermal emission from the original radio map.

\end{document}